\documentclass[
aps,%
12pt,%
final,%
notitlepage,%
oneside,%
onecolumn,%
nobibnotes,%
nofootinbib,%
superscriptaddress,%
noshowpacs,%
centertags]%
{revtex4}

\usepackage{lscape}
\usepackage{multirow}
\usepackage{longtable}
\usepackage[pdftex]{graphicx}

\begin{document}

\pagenumbering{arabic}

\title{Atlas and wavenumber tables for visible part of the rovibronic multiline 
emission spectrum of the $D_2$ molecule. II. Wavenumber range $18161 \div 14379$ cm$^{-1}$ 
measured with moderate resolution.}

\author{B.~P.~Lavrov}
\email{lavrov@pobox.spbu.ru}
\author{I.~S.~Umrikhin}
\affiliation{
Faculty of Physics, St.-Petersburg State University, \\
St.-Petersburg, 198504, Russia}

\begin{abstract}
The visible part ($\approx 550 \div 696$ nm) of the 
multiline electronic-vibro-rotational emission spectrum of the $D_2$ molecule 
was recorded with moderate resolution (line widths $\approx 0.013$ nm). 
The resolution was limited by Doppler broadening of 
spectral lines. After numerical deconvolution of the recorded intensity 
distributions and proper calibration of the spectrometer the new set of 
wavenumber values was obtained. The results are reported in the form of an 
atlas divided into 43 sections covering about 1.5 nm, containing pictures of 
images in the focal plane of the spectrometer, intensity distributions in 
linear and logarithmic scales and the table containing wavenumber and relative 
intensity values for 5445 spectral lines together with existing line assignments.
\end{abstract}

\maketitle

The present paper represents the second part of the sequence of our work
entitled "Atlas and wavenumber tables for visible part of the rovibronic multiline 
emission spectrum of the $D_2$ molecule". The experimental technique and the
procedure of data processing have been already described in the part~I \cite{LU2012_A1}.

As it was already writen in the part~I
the visible emission spectrum of the $D_2$ plasma in the wavelength region
$\approx 419 \div 696$ nm was recorded and analyzed.
In the atlas and tables of the present work we report part of this spectrum,
namely the wavelength region ($\approx 550 \div 696$ nm).
It contains one line of the atomic deuterium ($D_\alpha$),
corresponding line of atomic hydrogen (impurity) and 5443 lines of molecular deuterium.
The results are reported in the form of the
atlas divided into 43 sections each covering about 1.5 nm, containing pictures of 
images in the focal plane of the spectrometer, intensity distributions in 
linear and logarithmic scales and the table containing wavenumber and relative 
intensity values for recognized spectral lines together with existing line assignments.
Positions of spectral lines obtained by the deconvolution are presented as "stick diagrams" 
indicating their wavenumbers and amplitudes.
The numbering of the lines (for every fifth line) is shown under the 
intensity distributions in linear scale.

For assigned triplet spectral lines all measured in the present work wavenumber values
together with the experimental data from
\cite{FSC1985, DabrHerz, Davies, DiekeBlue1935, Dieke1935_2, GloersenDieke, DiekePorto, FMZ1976}
were used for obtaining the set of optimal rovibronic energy levels using the method
of statistical analysis \cite{LRJetf2005}. 
Detailed description of the analysis will be provided elsewhere. 
The analysis is similar to that described in our previous work \cite{LU2008}, but
the observation of pseudo doublets \cite{LUZ2012} forced us to carry out 
the optimization in two stages. At the first stage spectral line wavenumber 
values for band systems having one common low electronic state $a^3\Sigma_g^+$ 
($n^3\Lambda_g \to a^3\Sigma_g^+$, with $\Lambda = 0, 1$ and $n = 3 - 9$) were
analyzed. Obtained values of rovibronic energy levels were fixed and then all
other wavenumber values were added to the optimization procedure. Such a 
two-stage procedure gave us opportunity to obtain 595 energy level values of 
$a^3\Sigma_g^+$, $n^3\Lambda_g$ with $\Lambda = 0, 1$ and $n = 3 - 9$
electronic states having small fine structure splitting value with high 
precision. The values for 450 energy level values of 
$c^3\Pi_u$, $n^3\Lambda_u$ with $\Lambda = 0, 1$ and $n = 3 - 9$
electronic states are less accurate due to observed spectral lines fine 
structure.
Our statistical analysis shows good agreement (in the framework of the 
Rydberg-Ritz principle) between wavenumbers of spectral lines spread over 
the very wide range of wavenumbers $0.896 \div 28166.84$ cm$^{-1}$ (from RF up to UV) 
obtained for various band systems, by various methods and 
authors, and in various works.

Table contains: first column --- spectral line number $K$, 
second and third column --- measured wavenumber $\nu$ and 
intensity $I$ values respectively with standard deviation 
in units of last significant digit, fourth column --- wavenumber value 
of the line from \cite{FSC1985} in the cases when it was used as a 
reference data and the fifth column --- assignment in the Dieke's notations.
Confirmed by statistical analysis assignments for triplet lines are 
shown in bold and the new assignments are shown in italic.

\begin{acknowledgments}
Present work was supported, in part, by 
The Russian Foundation for Basic Research, Grant No. 10-03-00571-a.
\end{acknowledgments}

\newpage

\begin{landscape}
{\def\baselinestretch{1.0}
\footnotesize		
\setlength{\tabcolsep}{1pt}

\begin{longtable}[]{r|lr|l|r|r|lr|l|r}
\caption[]{$D_2$ rovibronic spectral lines vacuum wavenumbers values, obtained in present work $\nu$.
The uncertainties of the $\nu$ value determination (one SD) are shown in brackets
in units of last significant digit.
$\nu_R$ --- wavenumber from \cite{FSC1985} used as reference data.}\label{tabnewlines01}
\endfirsthead

\multicolumn{10}{l}{Table~\ref{tabnewlines01} (Continued).}\\ 
\hline
K & $\nu$, cm$^{-1}$ & $I$, counts & $\nu_R$, cm$^{-1}$ & Assignment &
K & $\nu$, cm$^{-1}$ & $I$, counts & $\nu_R$, cm$^{-1}$ & Assignment \\
\hline
\endhead

\hline
K & $\nu$, cm$^{-1}$ & $I$, counts & $\nu_R$, cm$^{-1}$ & Assignment &
K & $\nu$, cm$^{-1}$ & $I$, counts & $\nu_R$, cm$^{-1}$ & Assignment \\
\hline

6546	& 18160.01(2)	& 82.1(5)
& 18160.00	& S+  3E-2B  (3-9) P3    
& 6571	& 18144.16(3)	& 26.3(6)
&      	&  
\\

6547	& 18159.24(3)	& 13.1(5)
&      	&  
& 6572	& 18143.01(3)	& 17.6(4)
&      	& S-  4E-2C  (2-2) Q3    
\\

6548	& 18158.06(2)	& 55.1(5)
& 18158.07	&  
& 6573	& 18142.38(2)	& 51.8(4)
& 18142.39	&  
\\

6549	& 18157.07(3)	& 29.2(6)
&      	&  
& 6574	& \textbf{18141.83(3)}	& 83(4)
&      	& T+  3d-2c  (2-1) R1    
\\

6550	& 18156.58(3)	& 43.4(6)
& 18156.58	&  
& & & &	& S+  GK-2B  (6-8) P2    
\\

6551	& 18156.11(3)	& 27.5(8)
&      	&
& 6575	& 18141.60(3)	& 76(3)
&      	&
\\

6552	& 18155.71(2)	& 218.0(11)
& 18155.70	&  
& 6576	& 18141.21(3)	& 18.8(8)
&      	&  
\\
\hline
\end{longtable}
}

\newpage

\end{landscape}

\newpage
\begin{figure}[!ht]
\includegraphics[angle=90, totalheight=0.9\textheight]{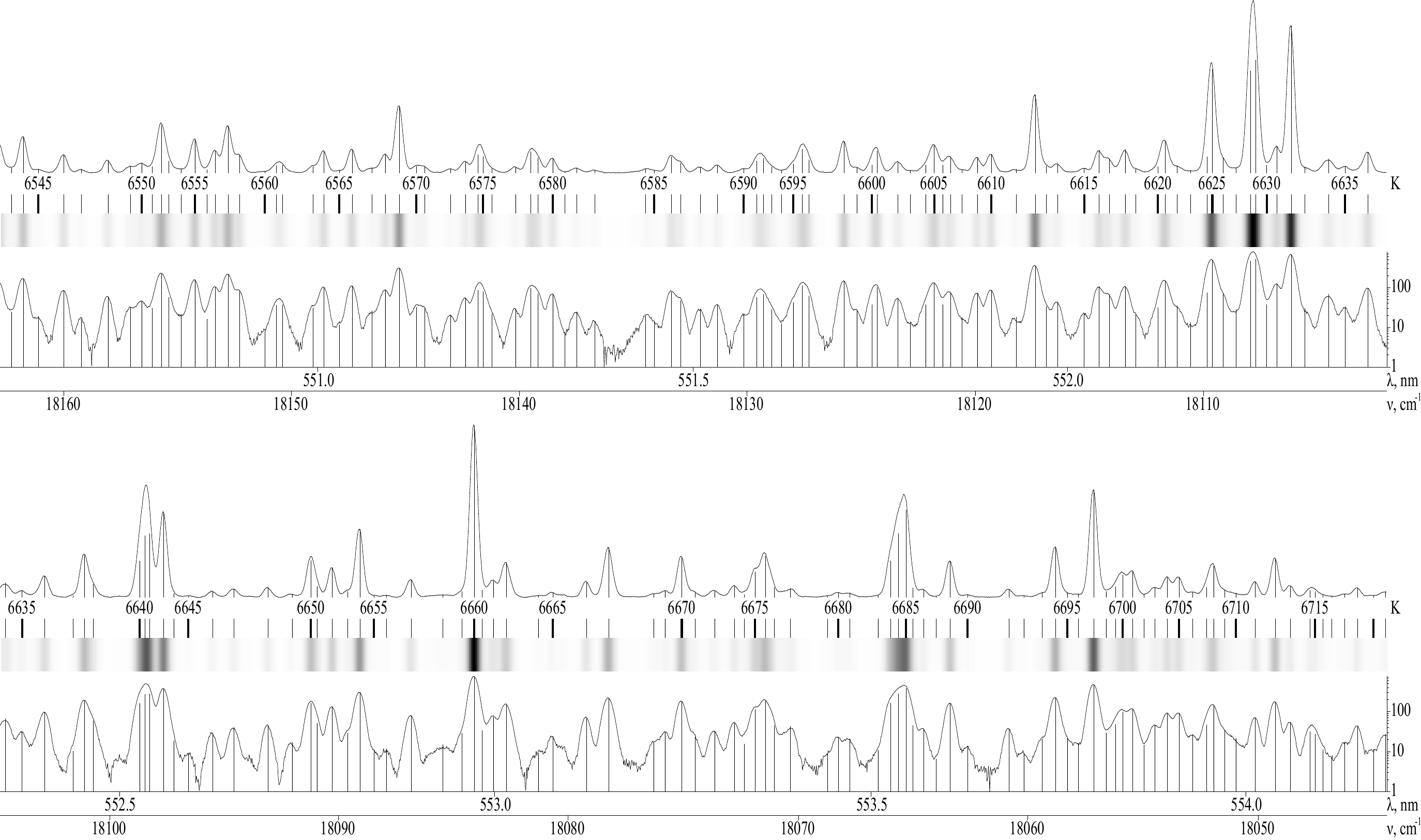}
\end{figure}

\newpage
\begin{figure}[!ht]
\includegraphics[angle=90, totalheight=0.9\textheight]{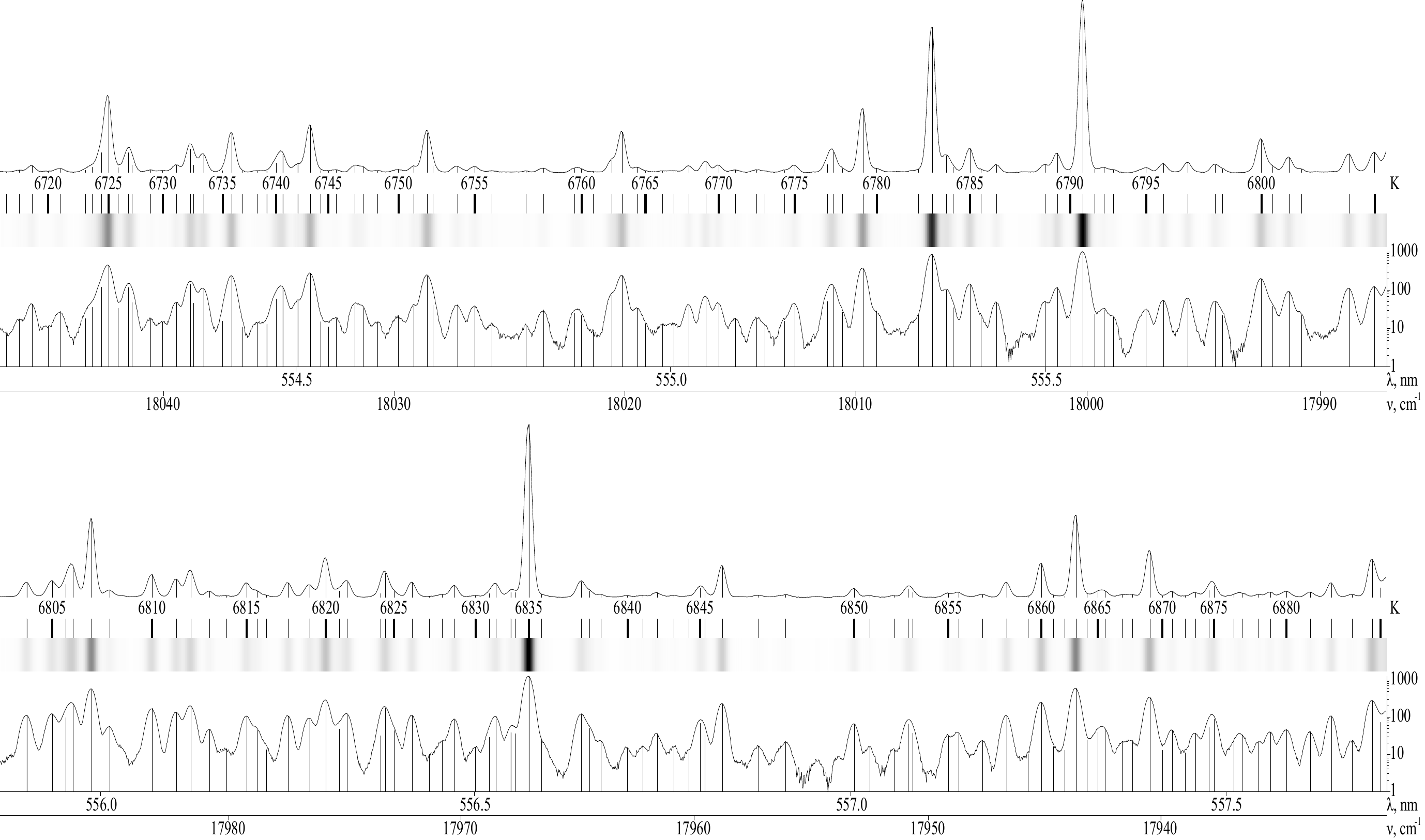}
\end{figure}

\newpage
\begin{figure}[!ht]
\includegraphics[angle=90, totalheight=0.9\textheight]{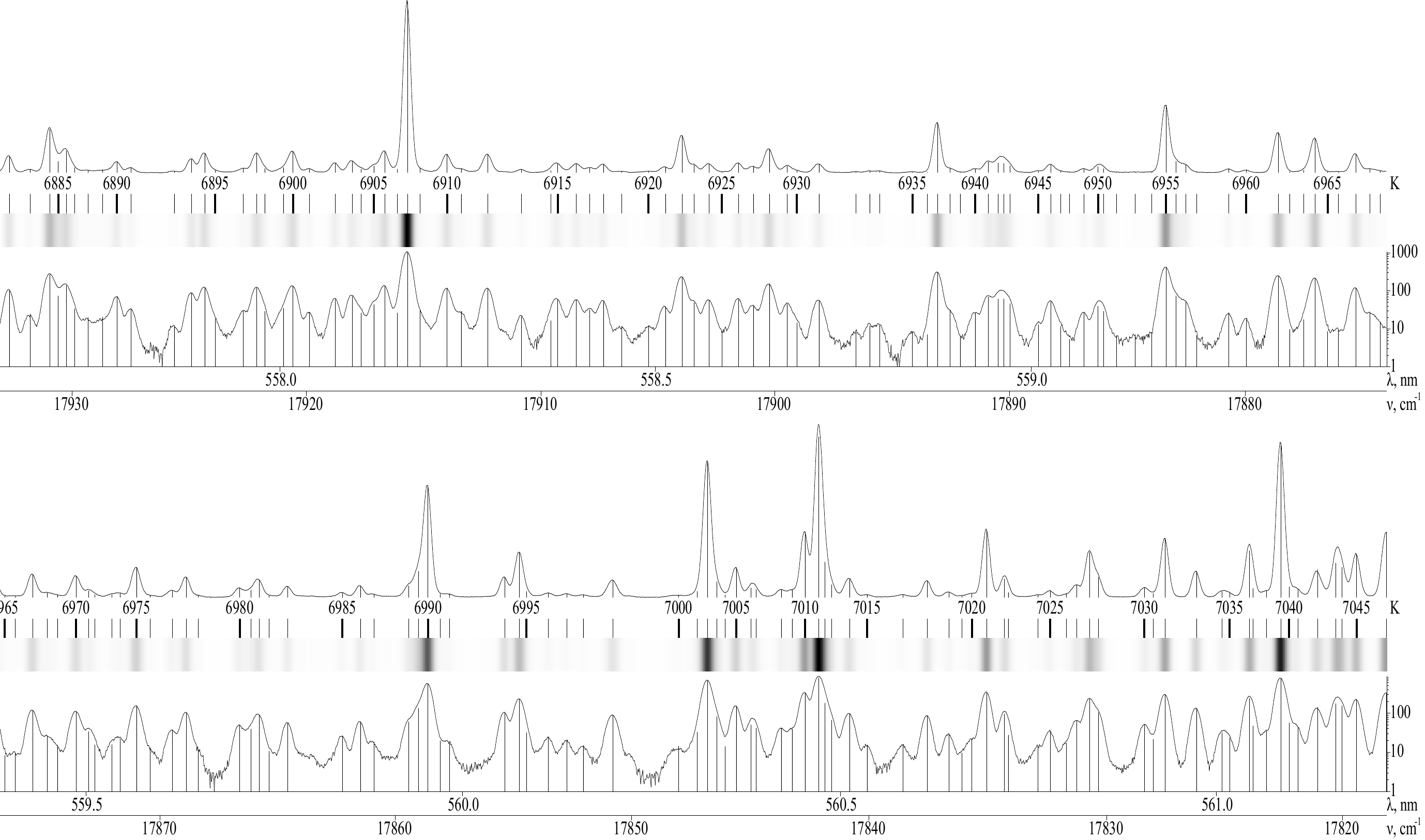}
\end{figure}

\newpage
\begin{figure}[!ht]
\includegraphics[angle=90, totalheight=0.9\textheight]{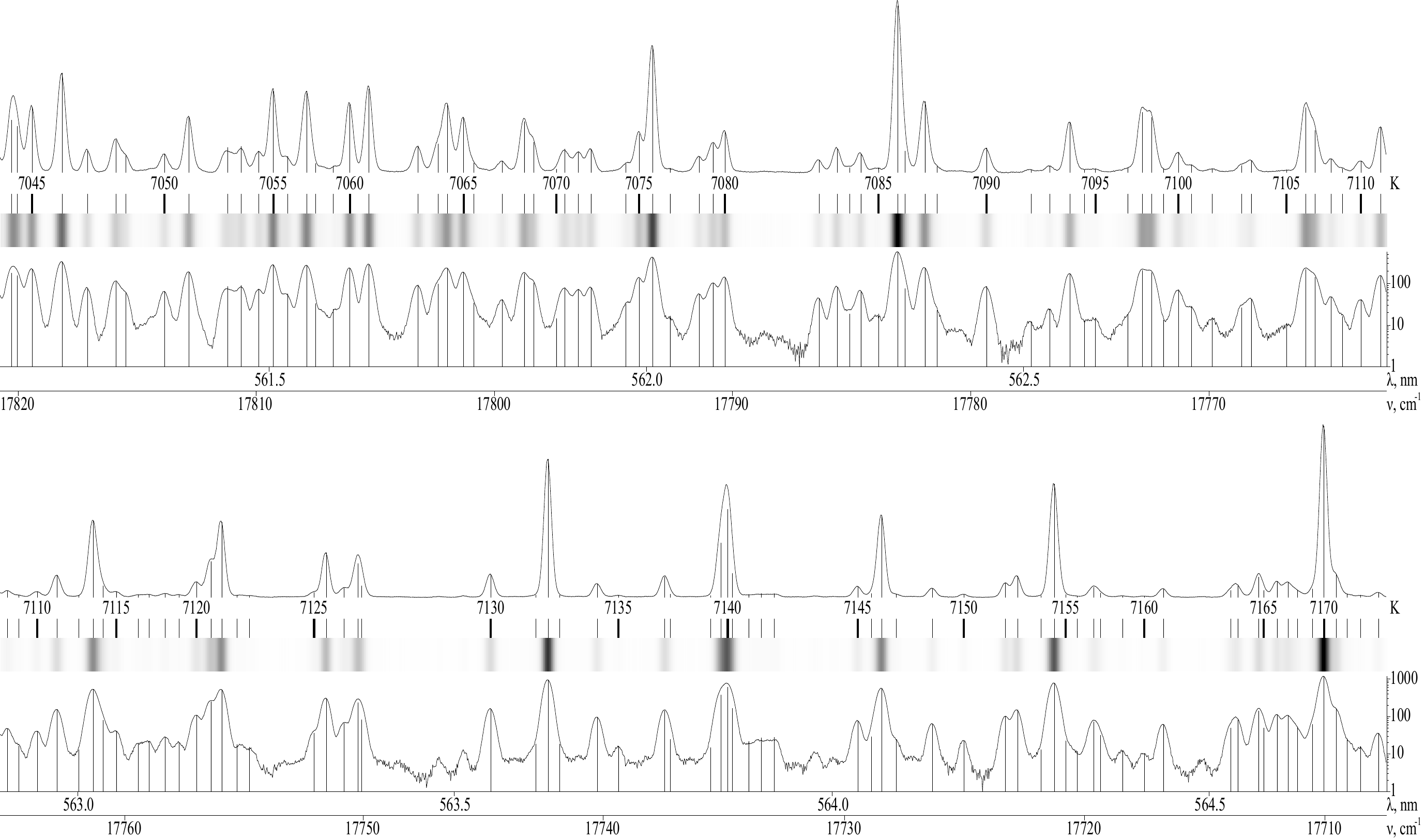}
\end{figure}

\newpage
\begin{figure}[!ht]
\includegraphics[angle=90, totalheight=0.9\textheight]{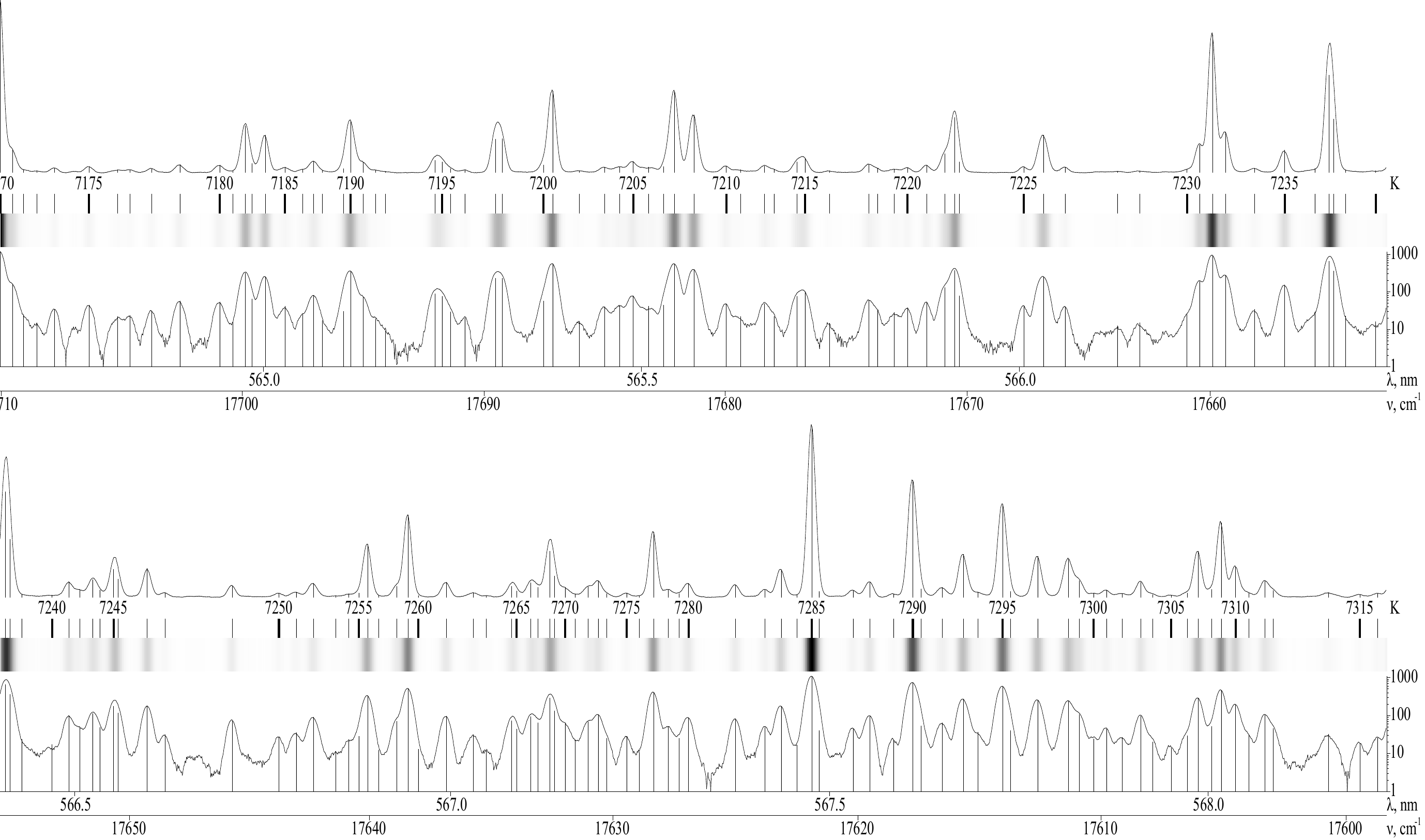}
\end{figure}

\newpage
\begin{figure}[!ht]
\includegraphics[angle=90, totalheight=0.9\textheight]{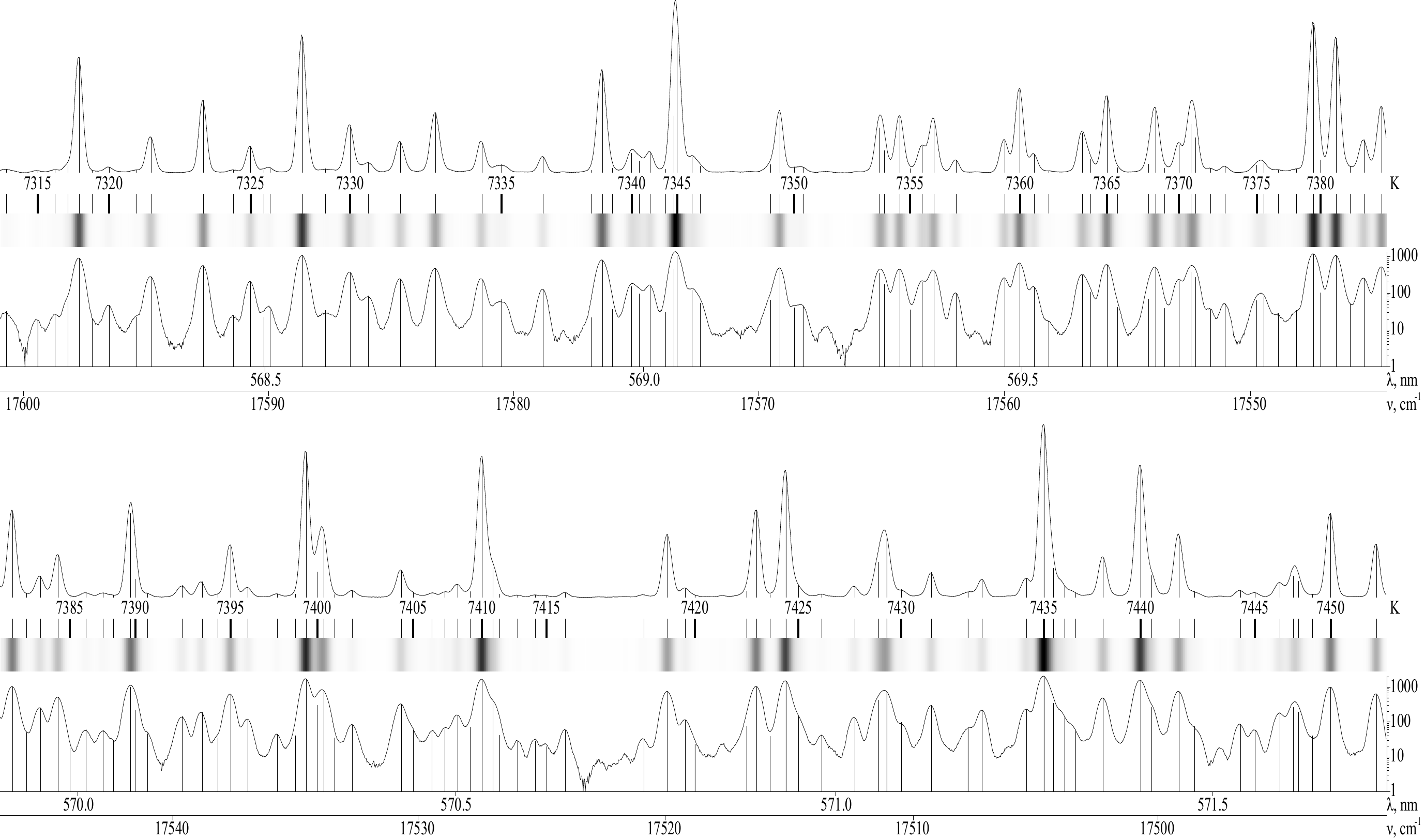}
\end{figure}

\newpage
\begin{figure}[!ht]
\includegraphics[angle=90, totalheight=0.9\textheight]{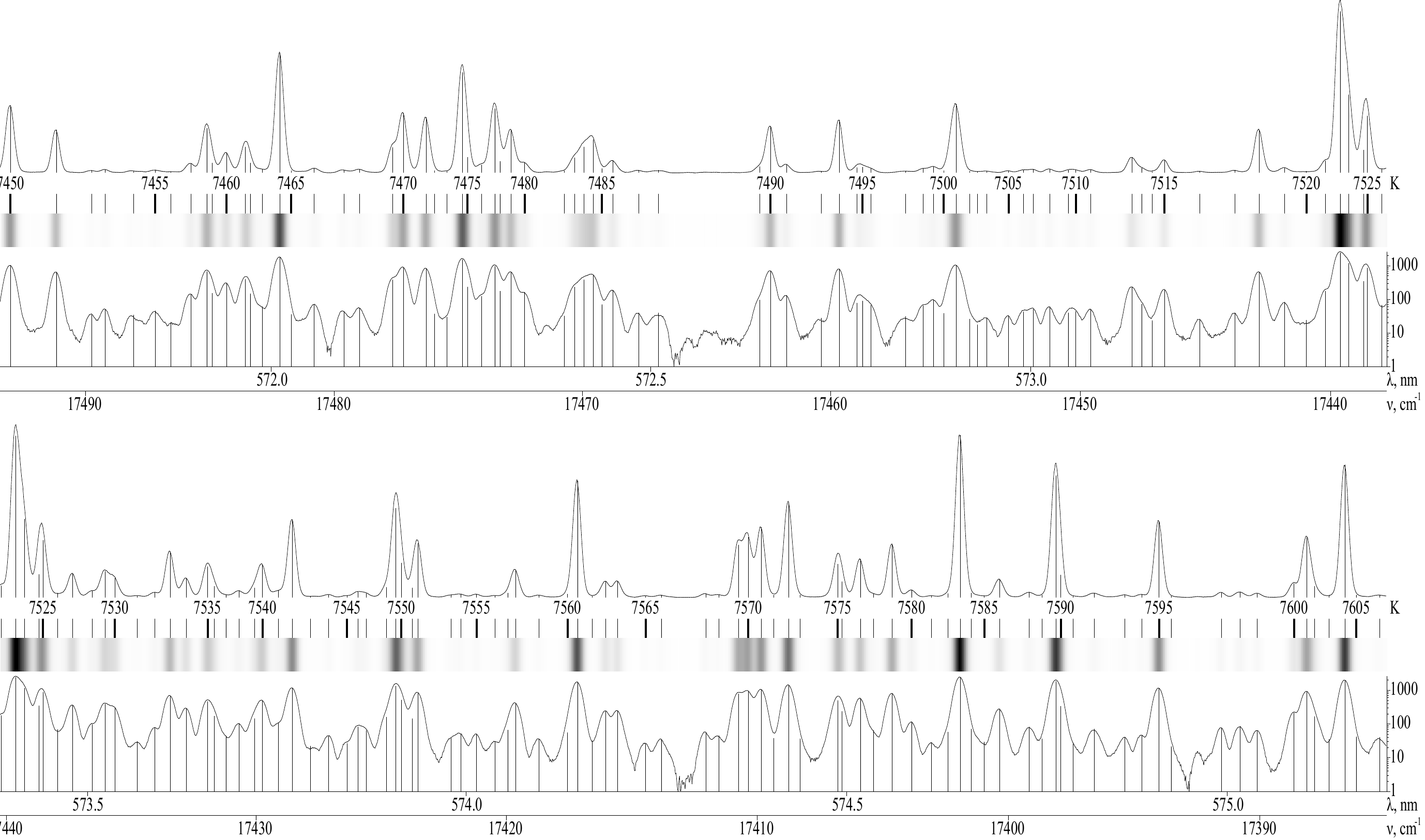}
\end{figure}

\newpage
\begin{figure}[!ht]
\includegraphics[angle=90, totalheight=0.9\textheight]{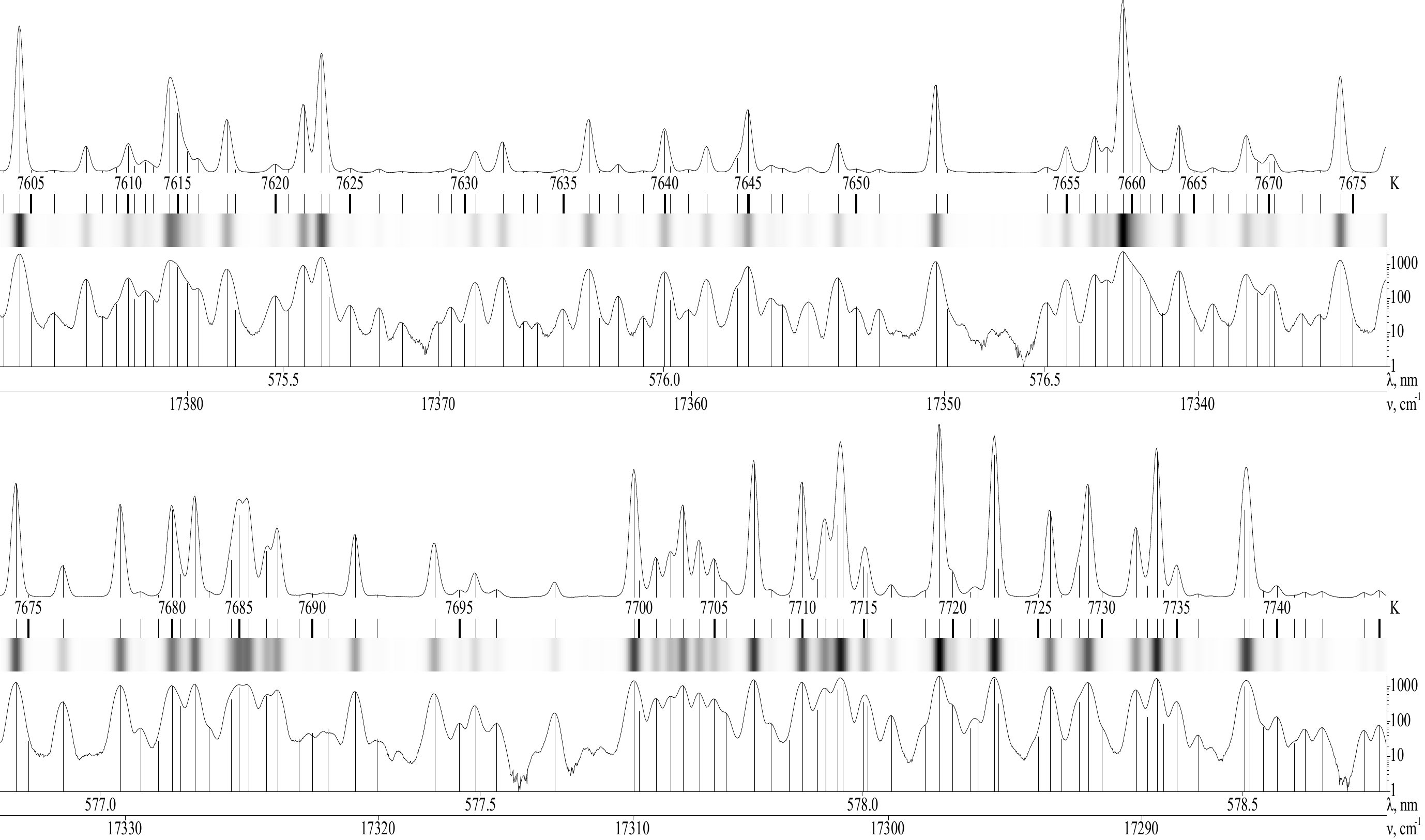}
\end{figure}

\newpage
\begin{figure}[!ht]
\includegraphics[angle=90, totalheight=0.9\textheight]{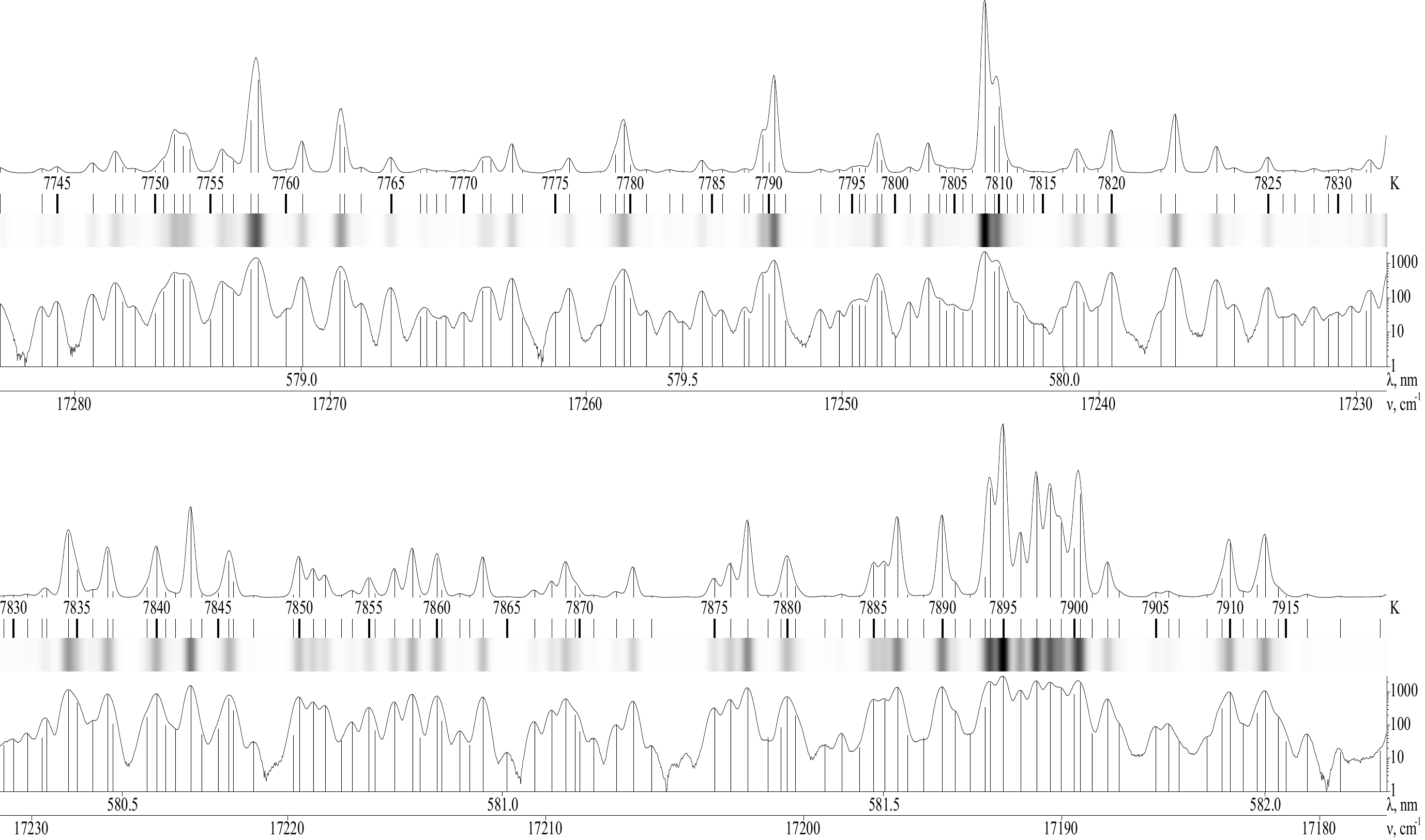}
\end{figure}

\newpage
\begin{figure}[!ht]
\includegraphics[angle=90, totalheight=0.9\textheight]{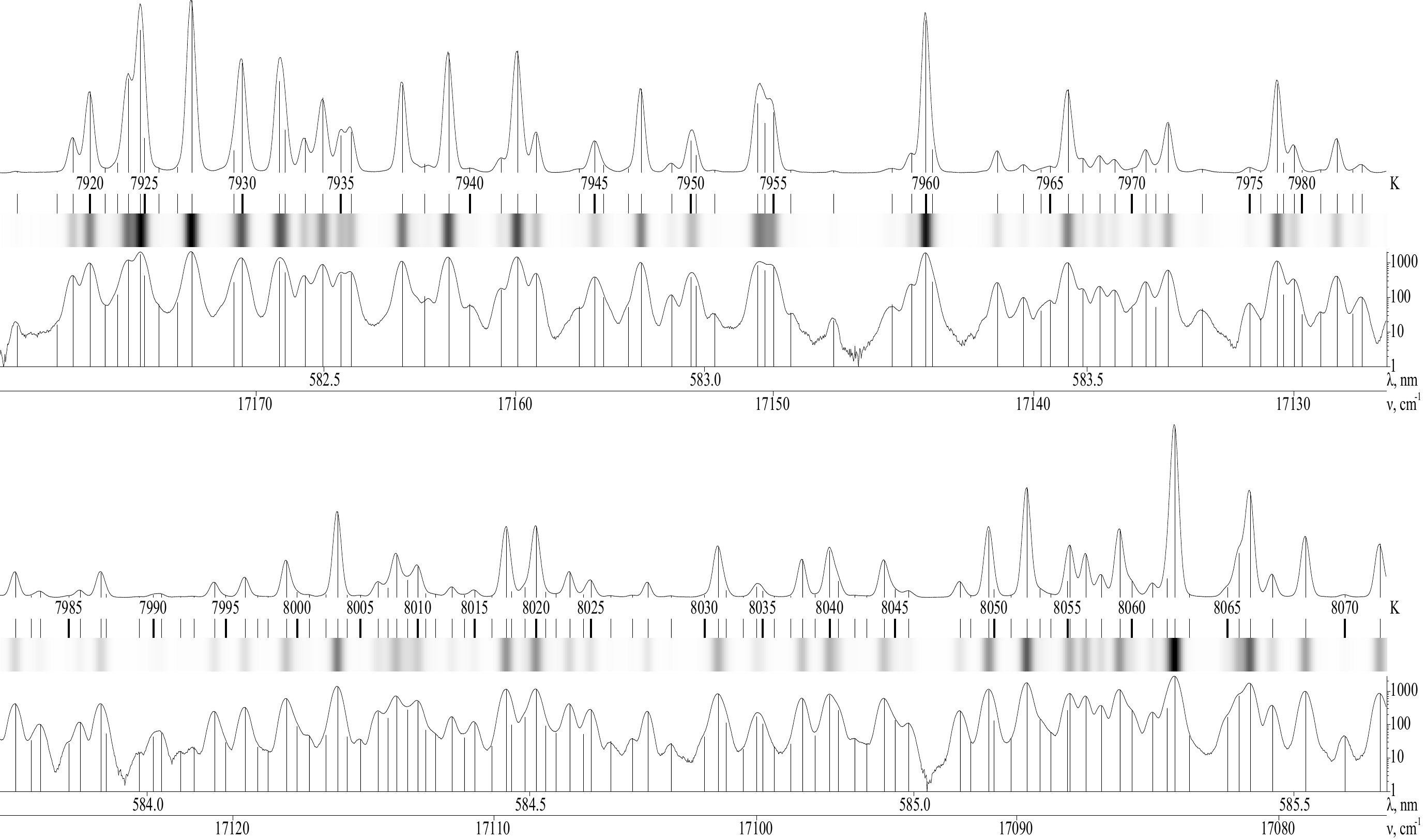}
\end{figure}

\newpage
\begin{figure}[!ht]
\includegraphics[angle=90, totalheight=0.9\textheight]{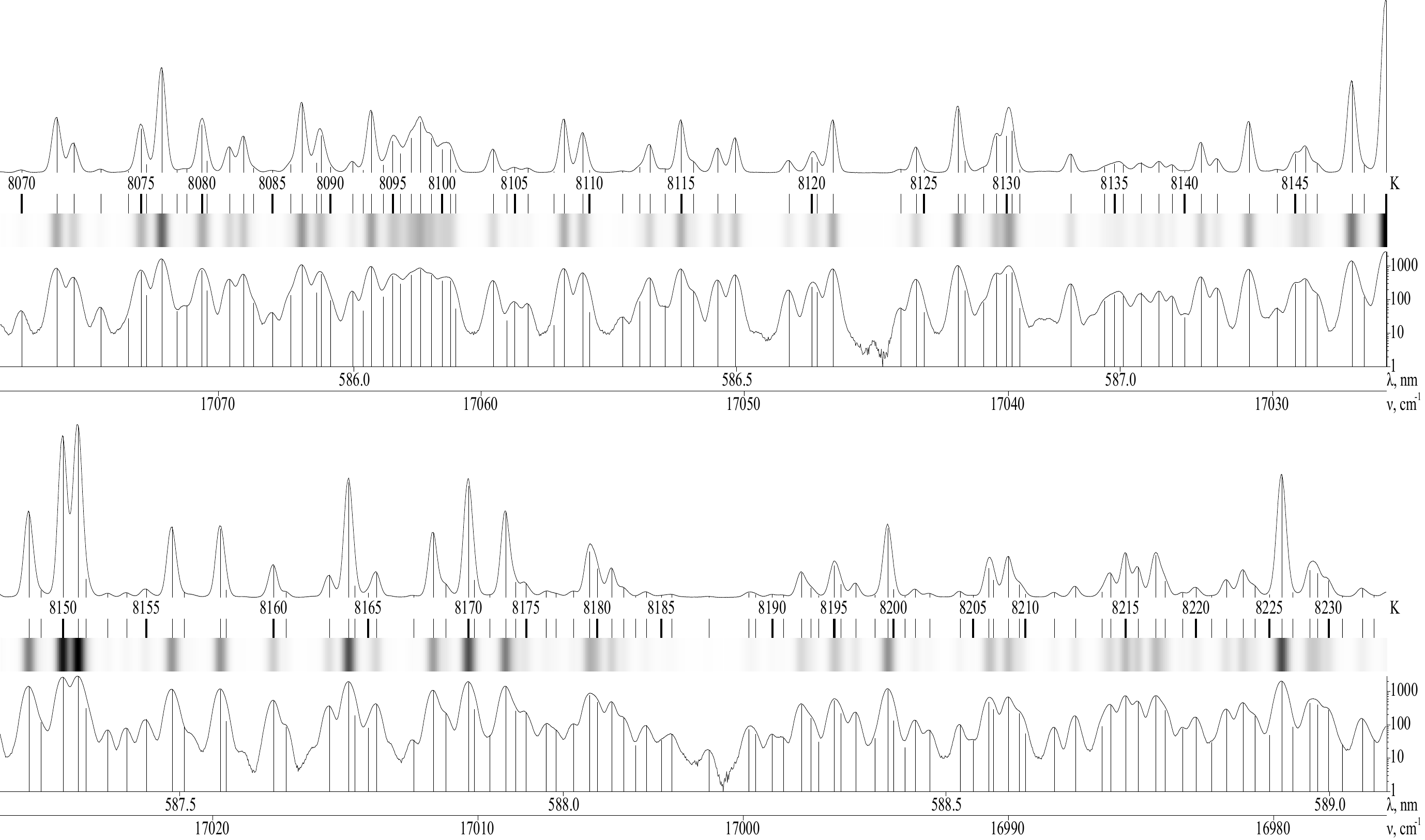}
\end{figure}

\newpage
\begin{figure}[!ht]
\includegraphics[angle=90, totalheight=0.9\textheight]{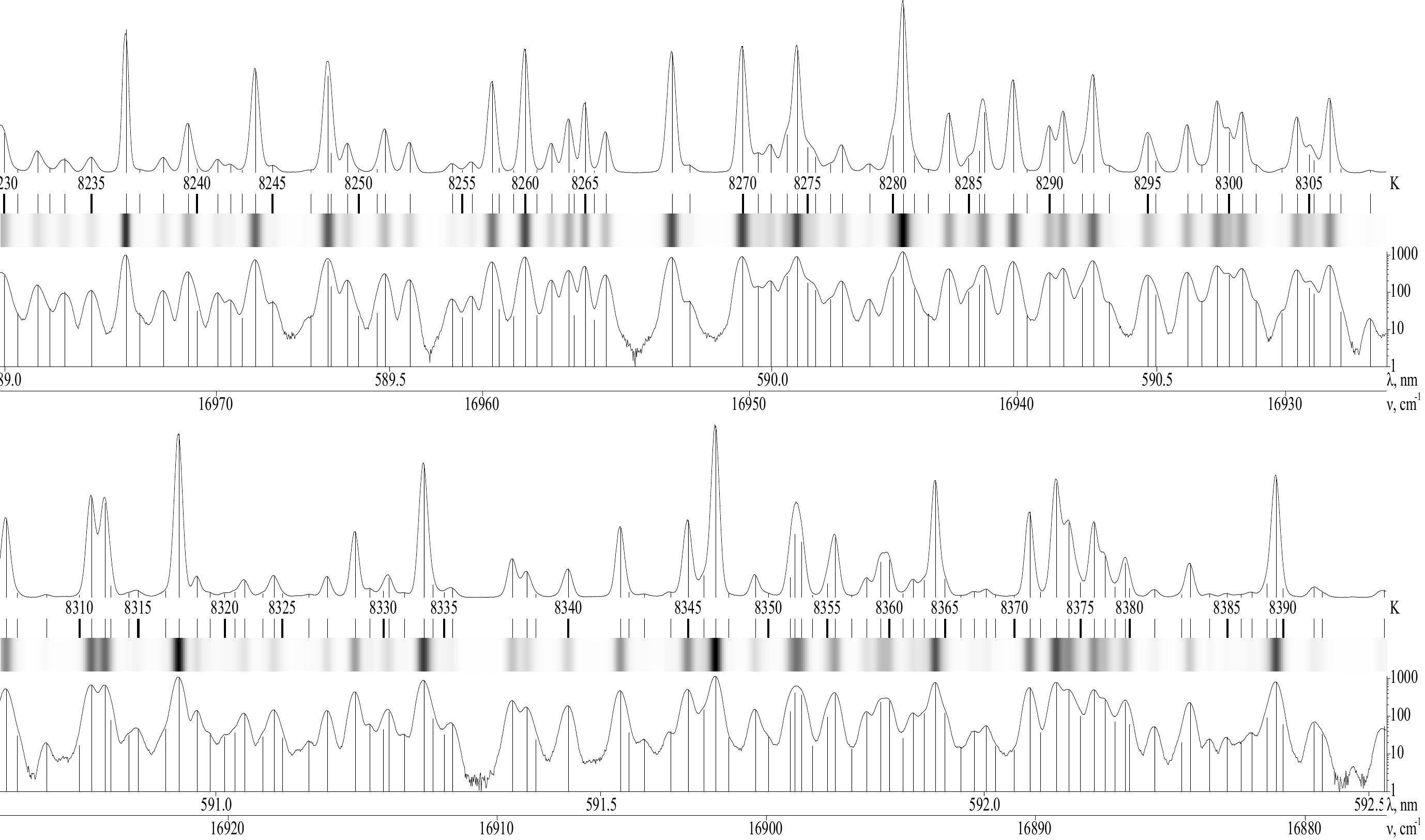}
\end{figure}

\newpage
\begin{figure}[!ht]
\includegraphics[angle=90, totalheight=0.9\textheight]{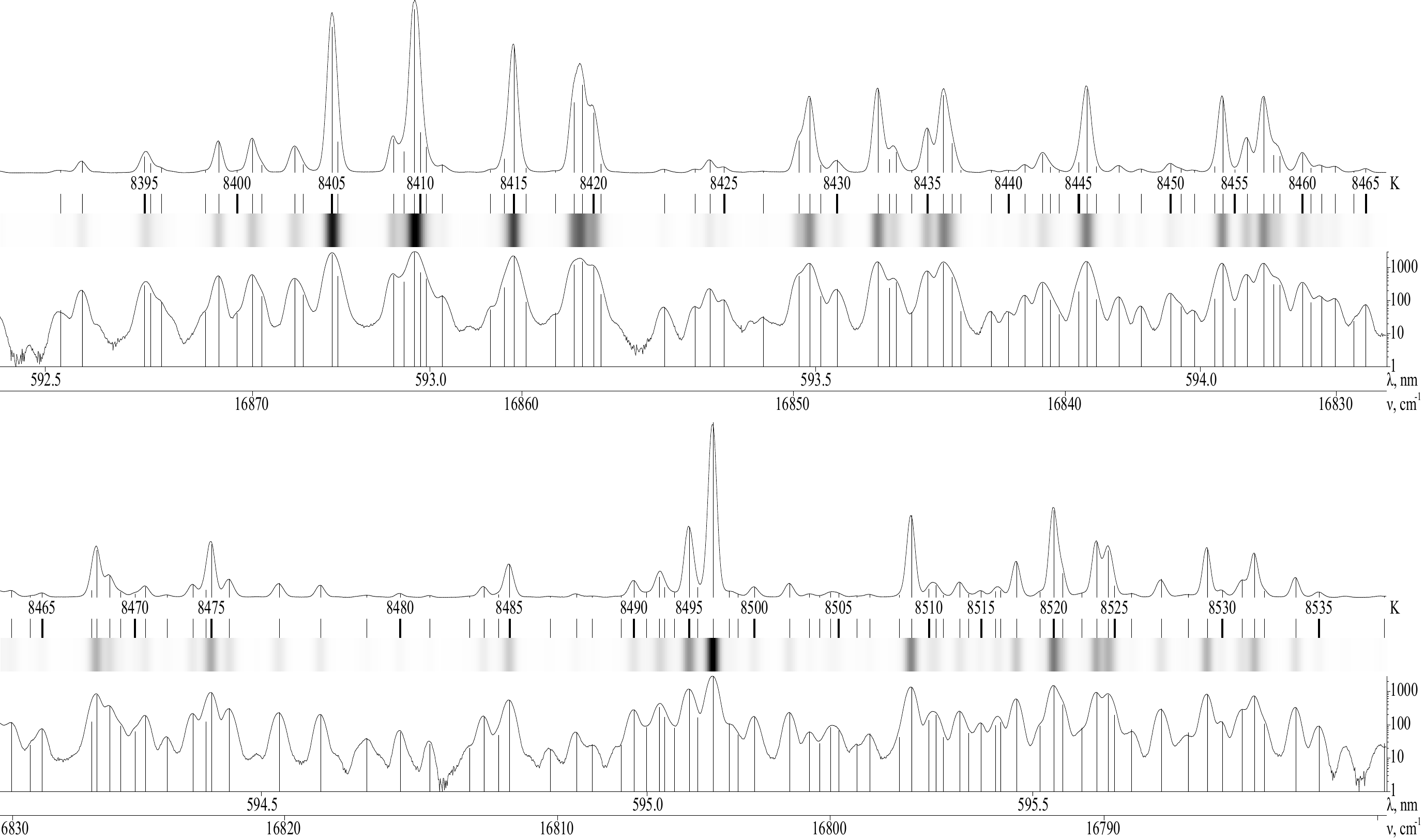}
\end{figure}

\newpage
\begin{figure}[!ht]
\includegraphics[angle=90, totalheight=0.9\textheight]{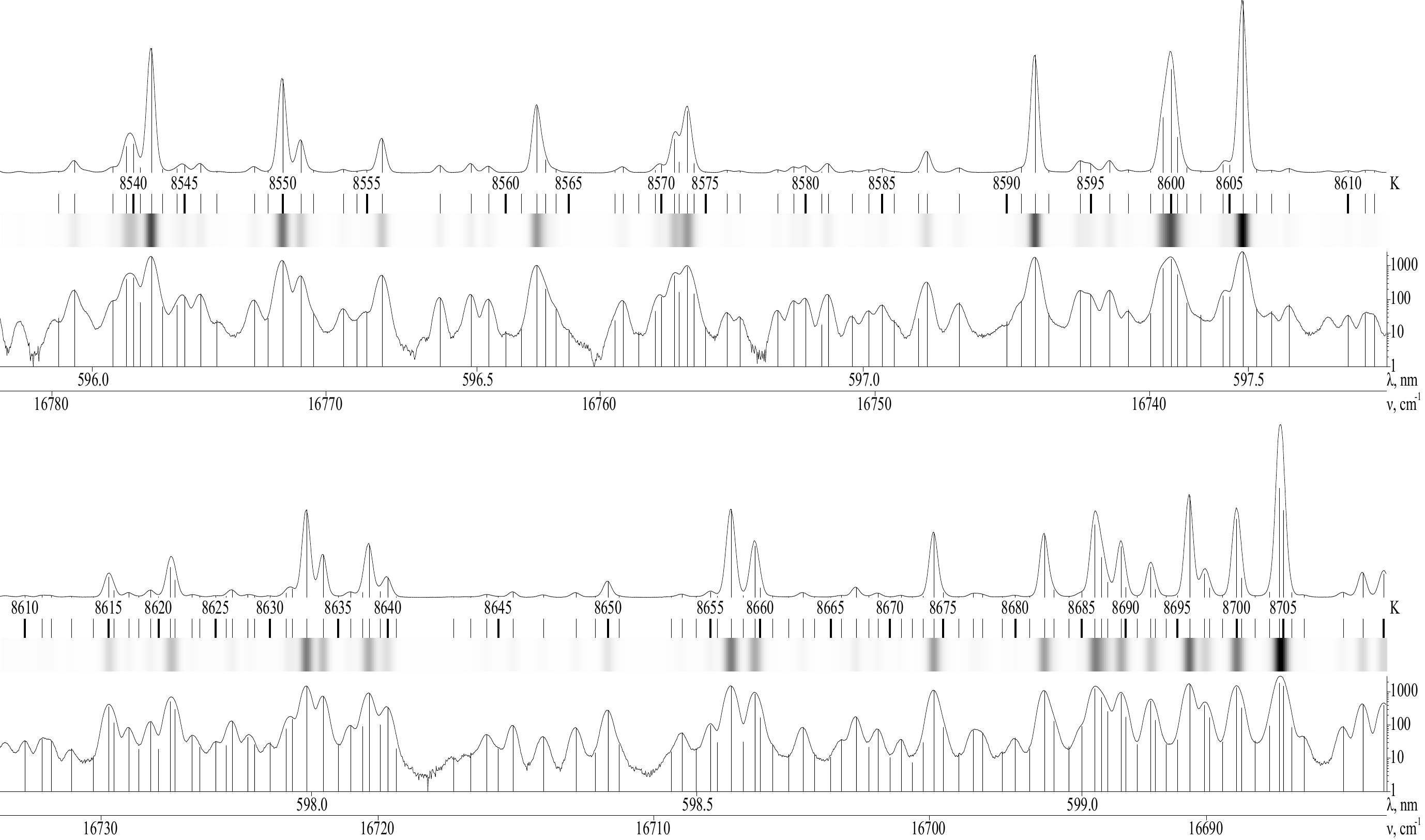}
\end{figure}

\newpage
\begin{figure}[!ht]
\includegraphics[angle=90, totalheight=0.9\textheight]{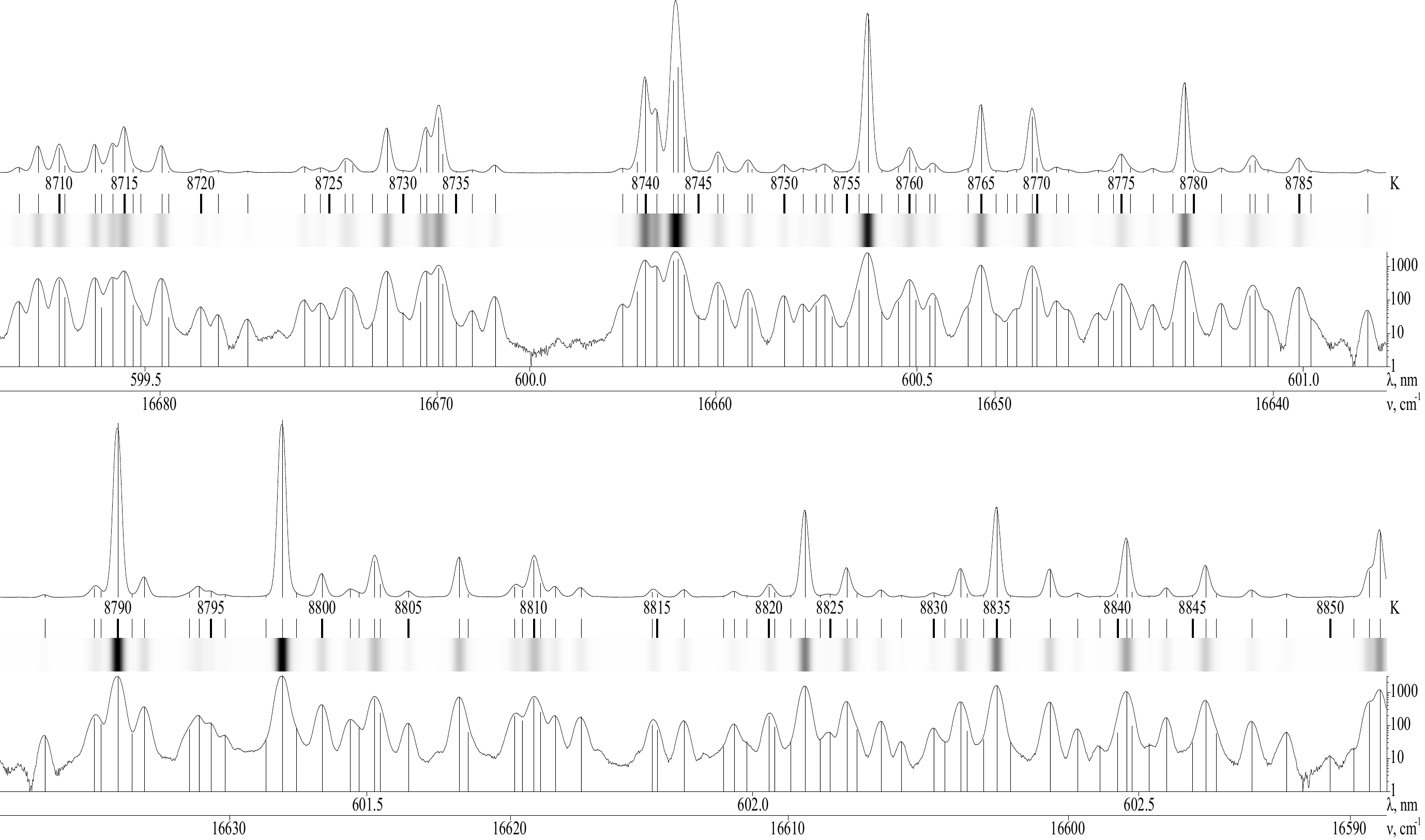}
\end{figure}

\newpage
\begin{figure}[!ht]
\includegraphics[angle=90, totalheight=0.9\textheight]{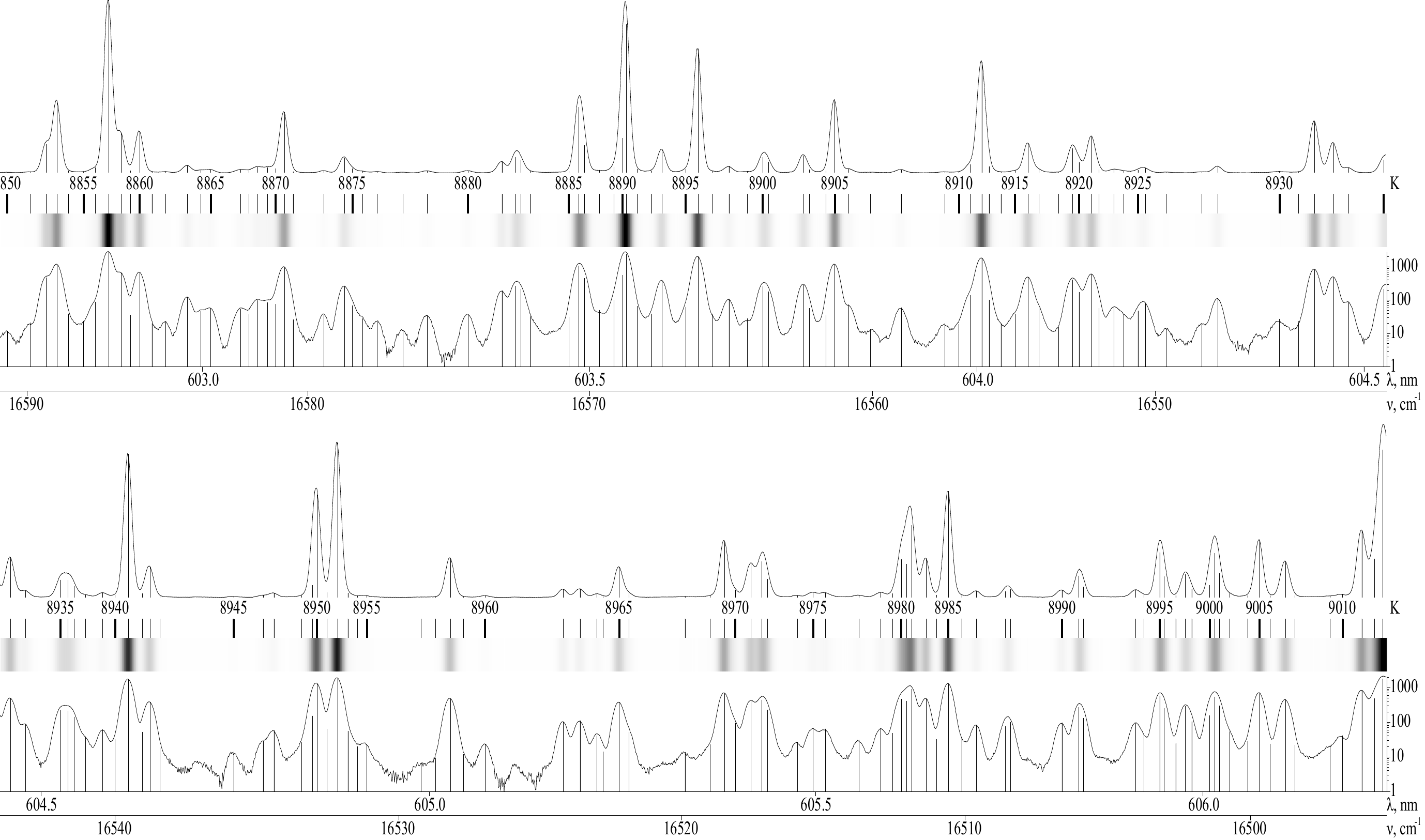}
\end{figure}

\newpage
\begin{figure}[!ht]
\includegraphics[angle=90, totalheight=0.9\textheight]{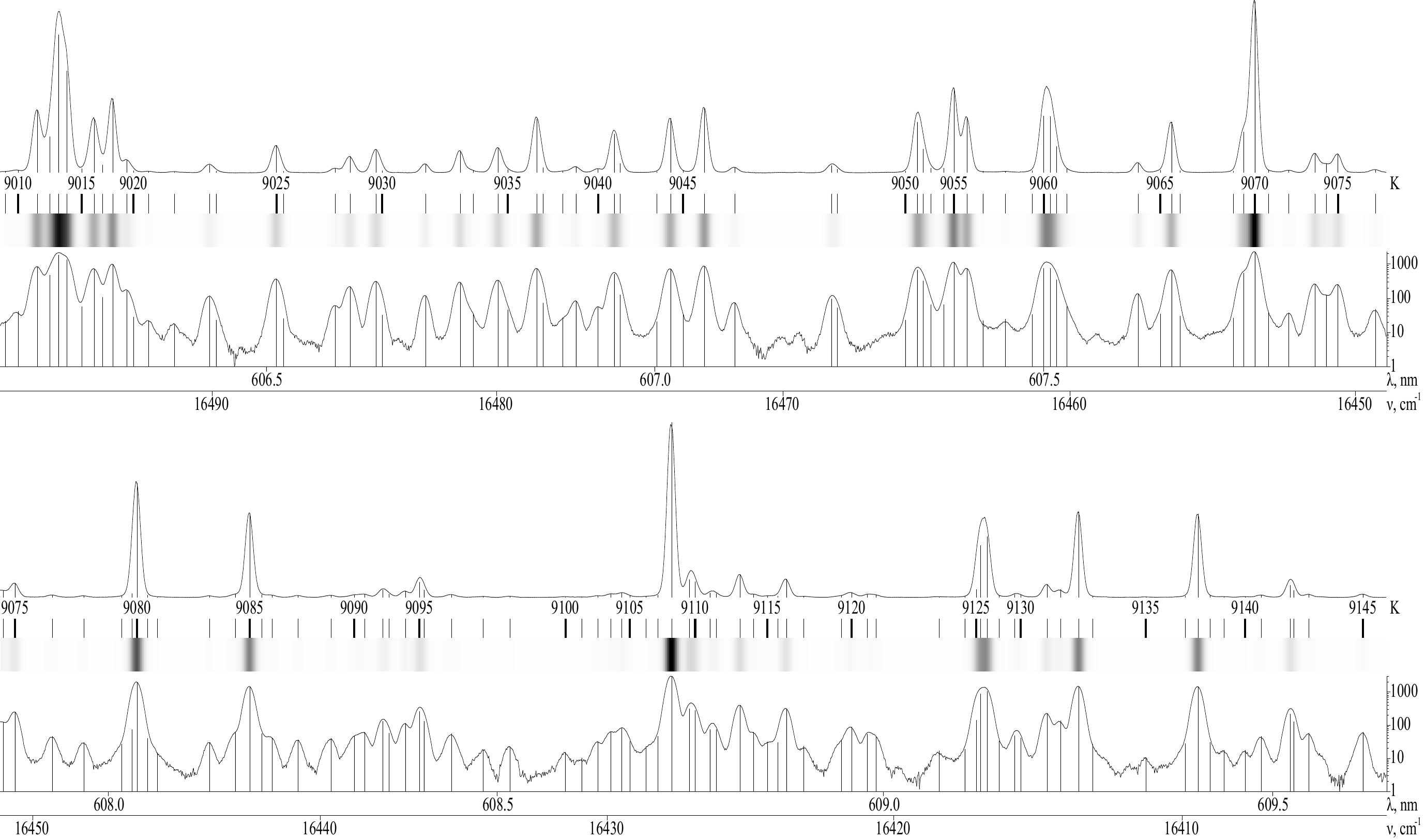}
\end{figure}

\newpage
\begin{figure}[!ht]
\includegraphics[angle=90, totalheight=0.9\textheight]{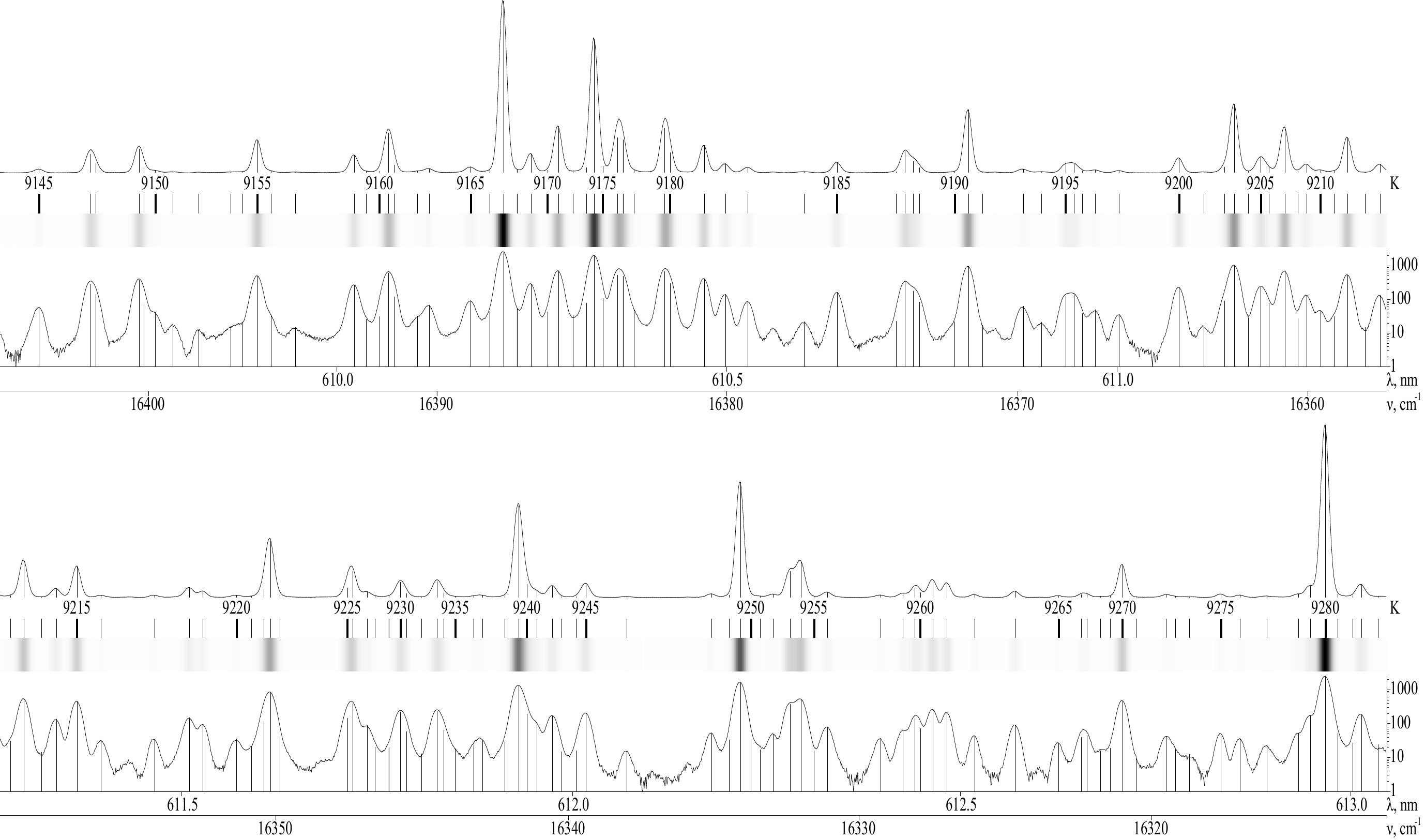}
\end{figure}

\newpage
\begin{figure}[!ht]
\includegraphics[angle=90, totalheight=0.9\textheight]{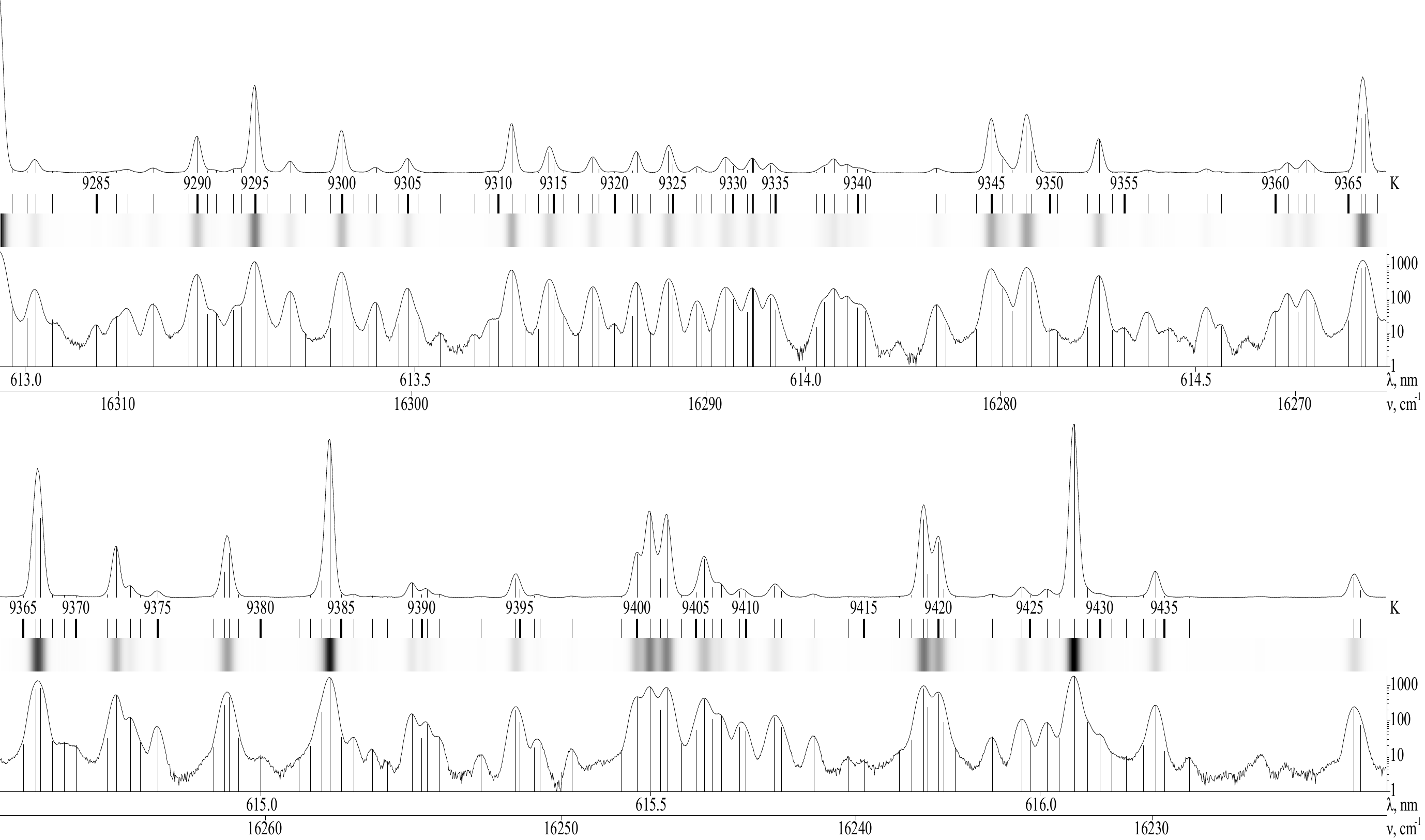}
\end{figure}

\newpage
\begin{figure}[!ht]
\includegraphics[angle=90, totalheight=0.9\textheight]{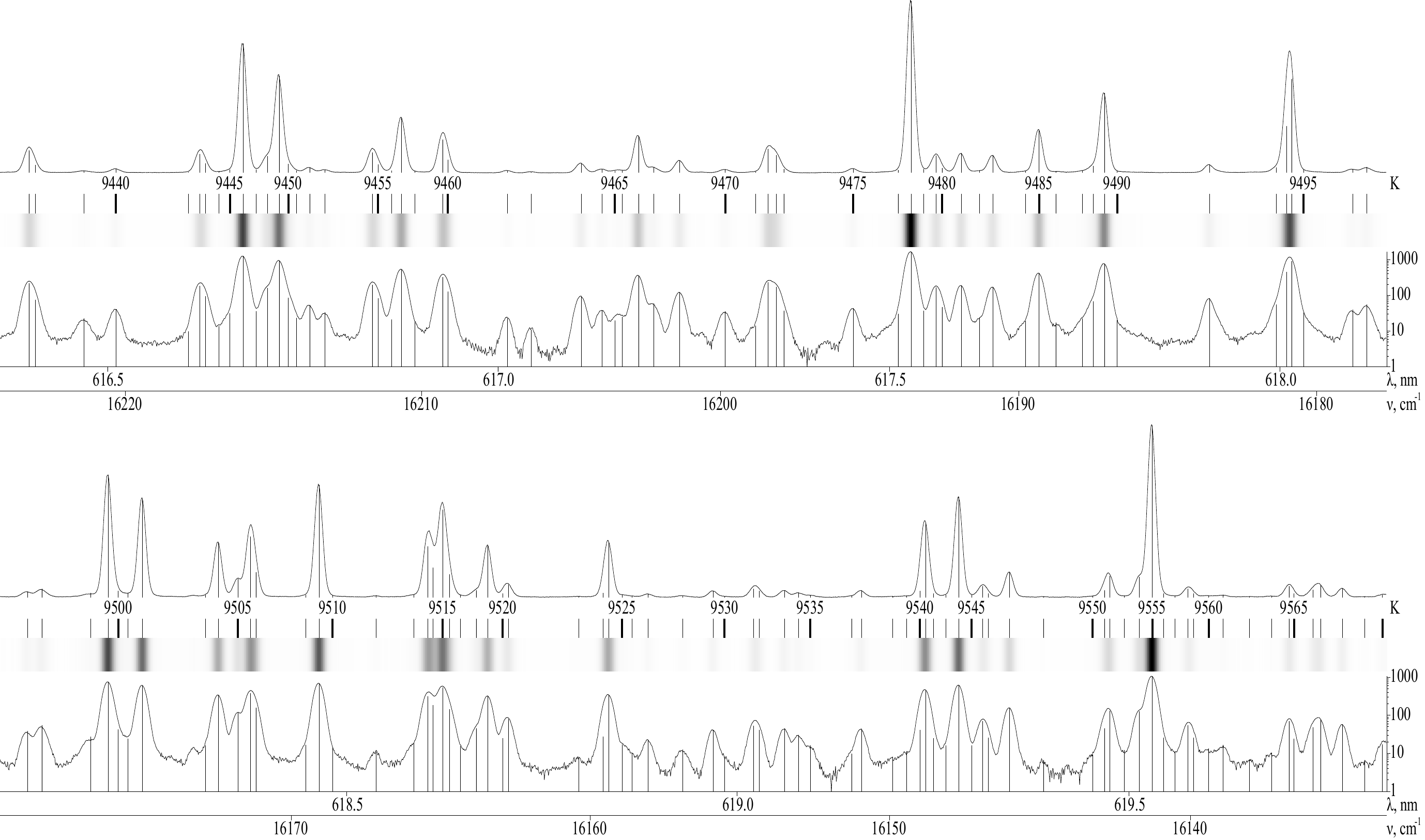}
\end{figure}

\newpage
\begin{figure}[!ht]
\includegraphics[angle=90, totalheight=0.9\textheight]{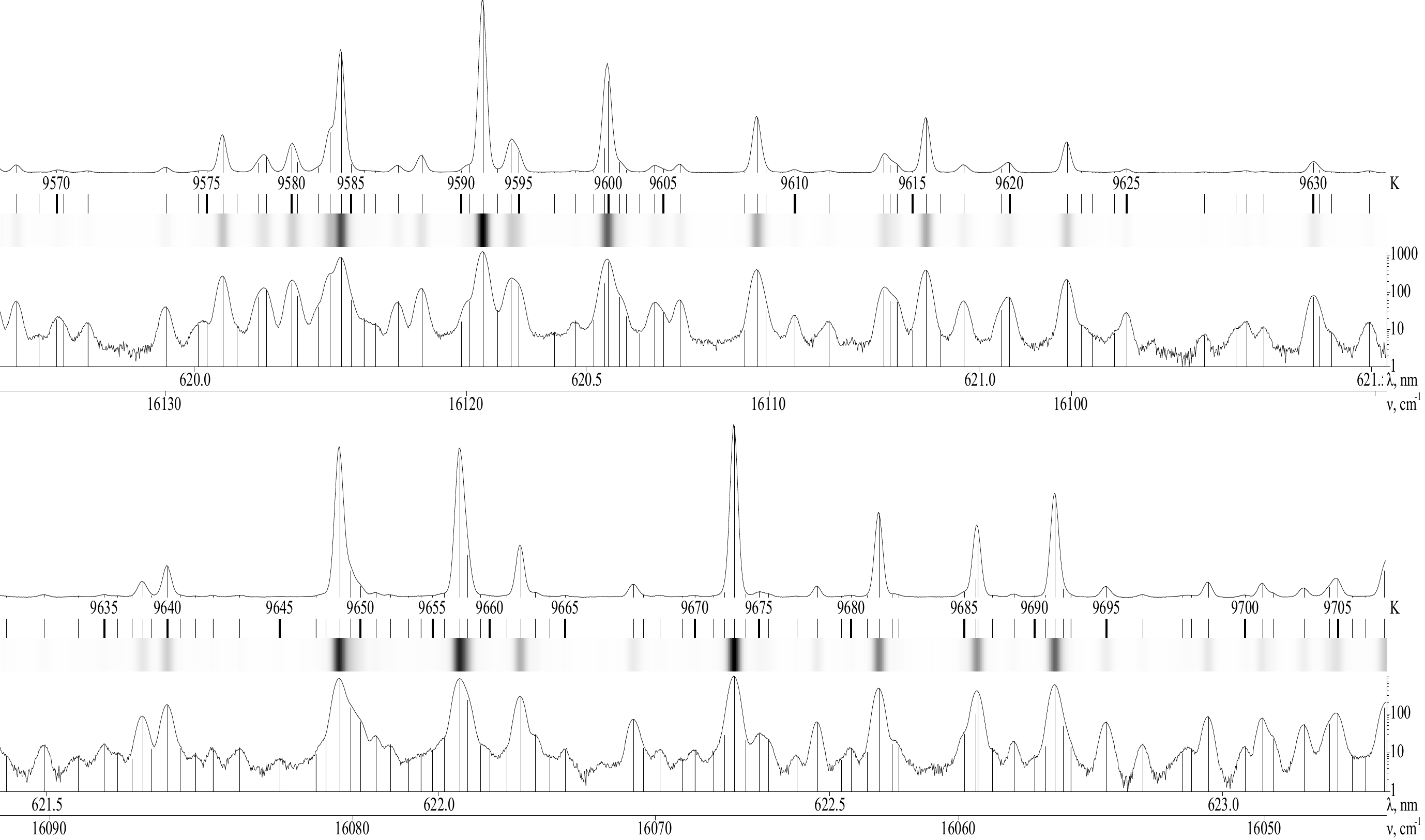}
\end{figure}

\newpage
\begin{figure}[!ht]
\includegraphics[angle=90, totalheight=0.9\textheight]{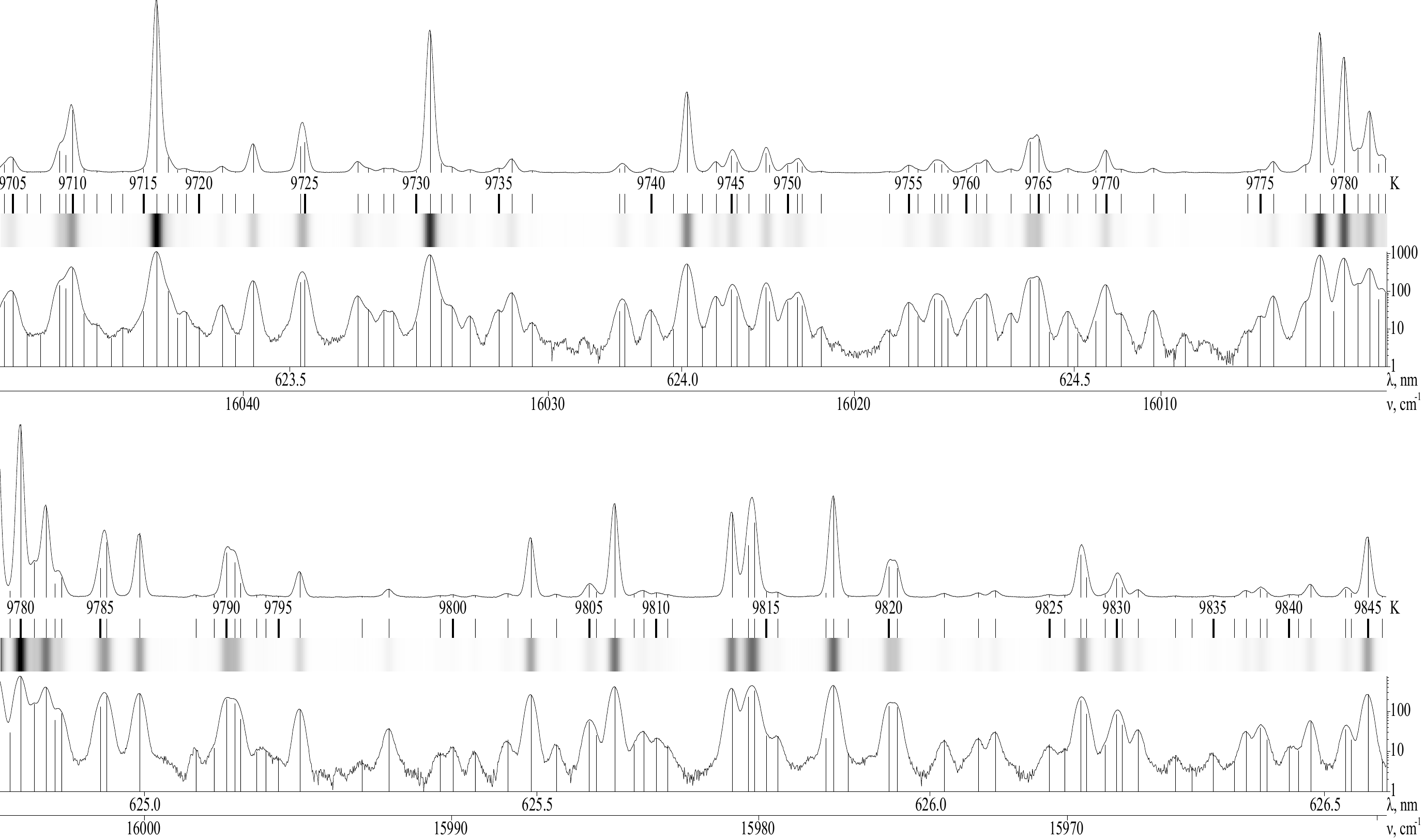}
\end{figure}

\newpage
\begin{figure}[!ht]
\includegraphics[angle=90, totalheight=0.9\textheight]{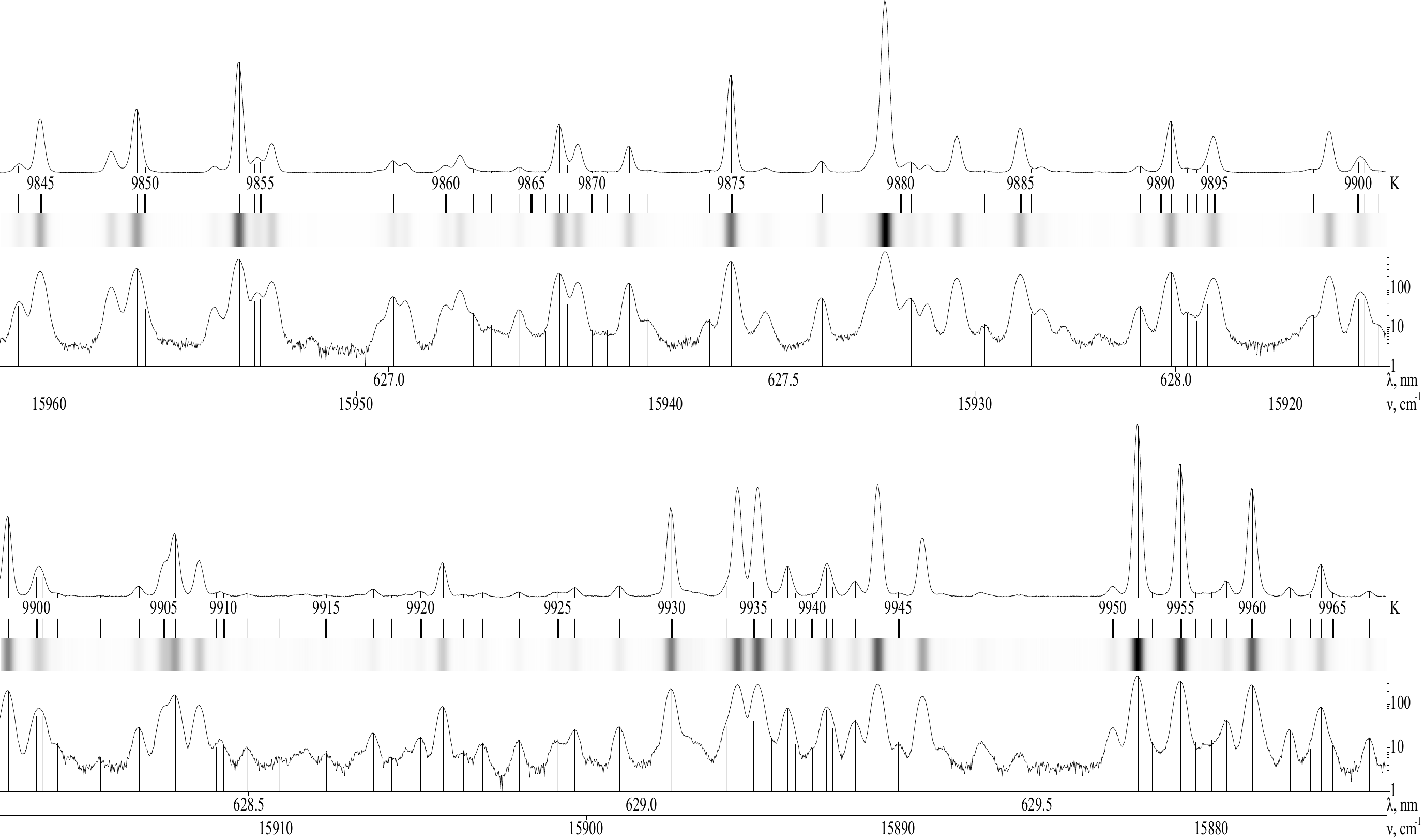}
\end{figure}

\newpage
\begin{figure}[!ht]
\includegraphics[angle=90, totalheight=0.9\textheight]{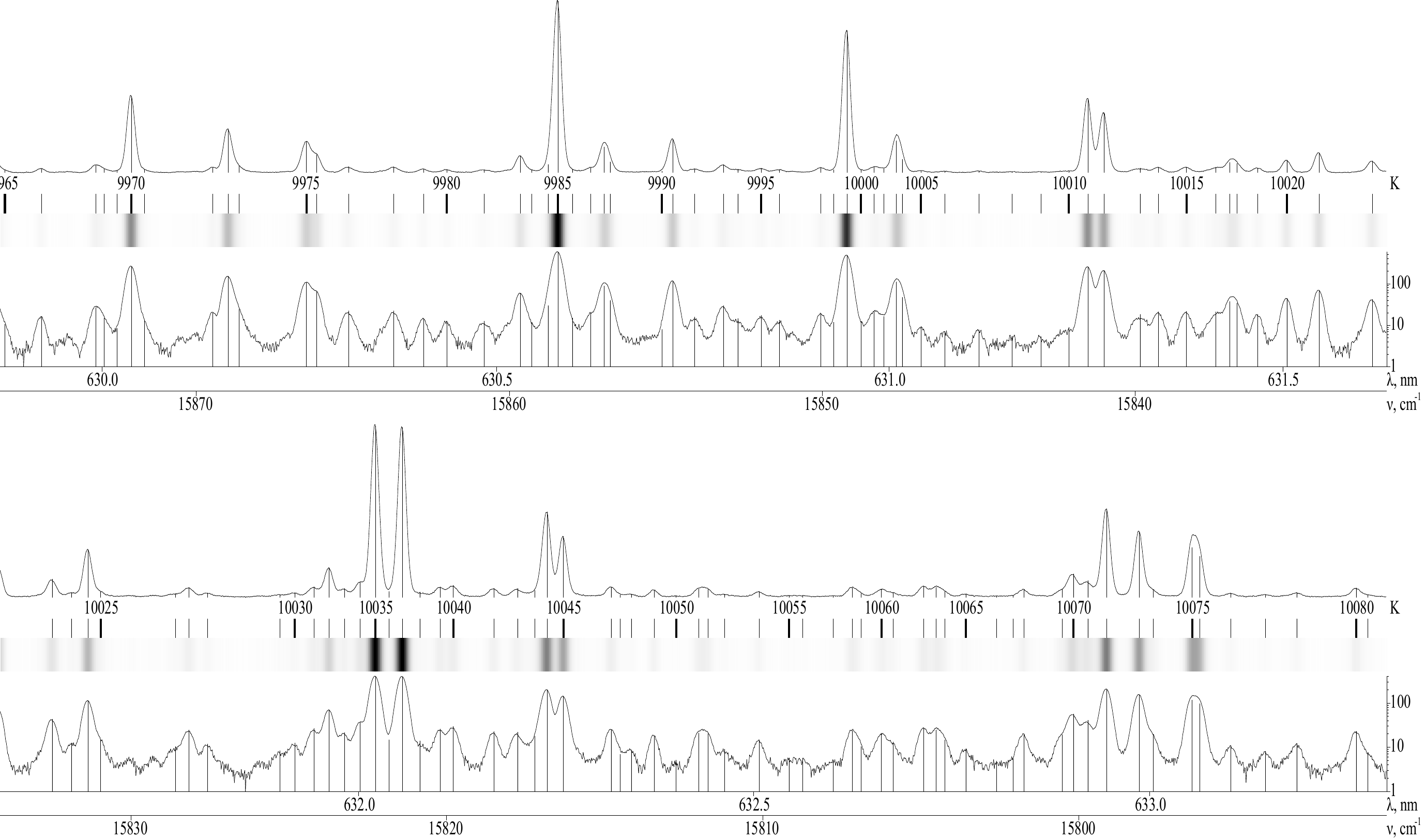}
\end{figure}

\newpage
\begin{figure}[!ht]
\includegraphics[angle=90, totalheight=0.9\textheight]{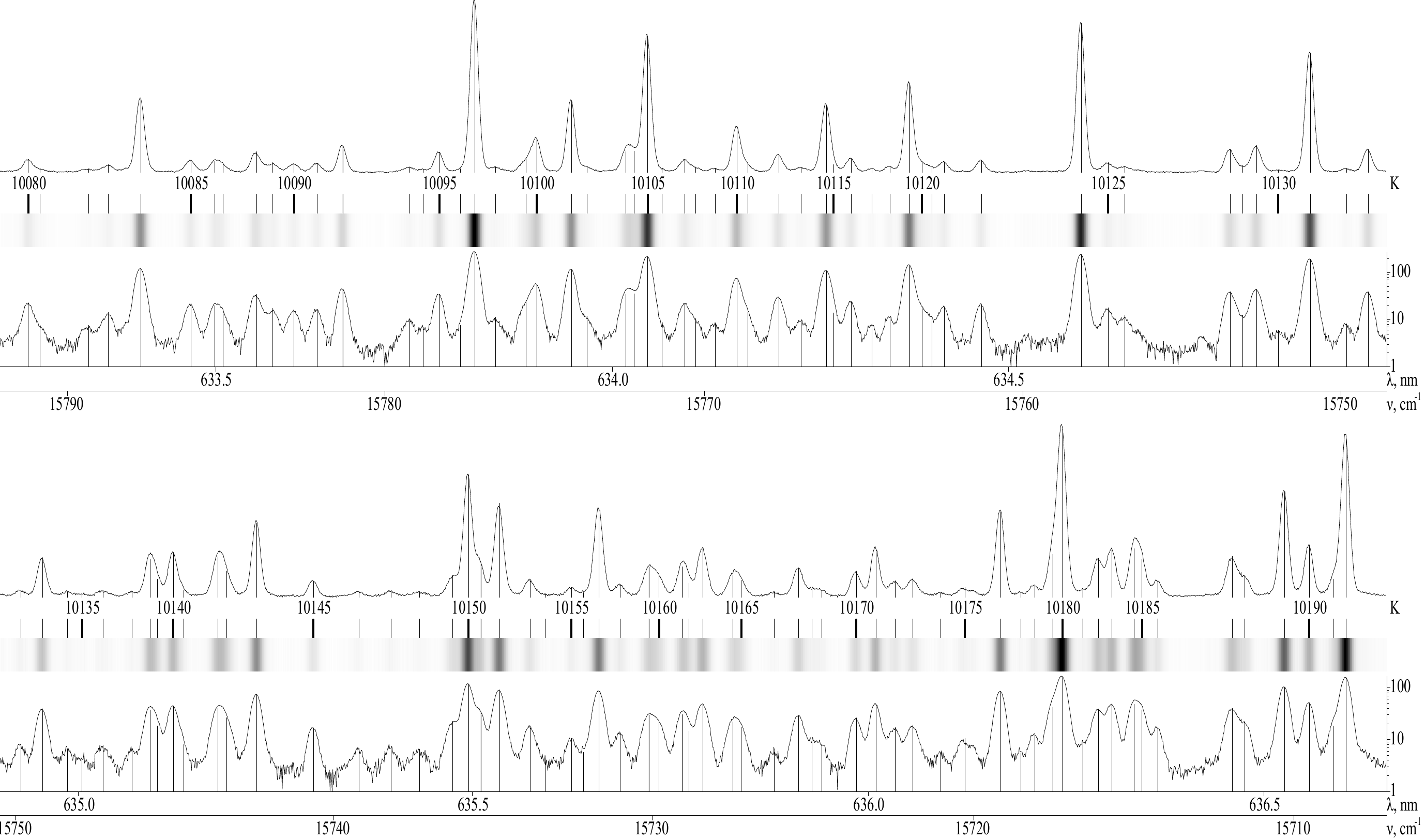}
\end{figure}

\newpage
\begin{figure}[!ht]
\includegraphics[angle=90, totalheight=0.9\textheight]{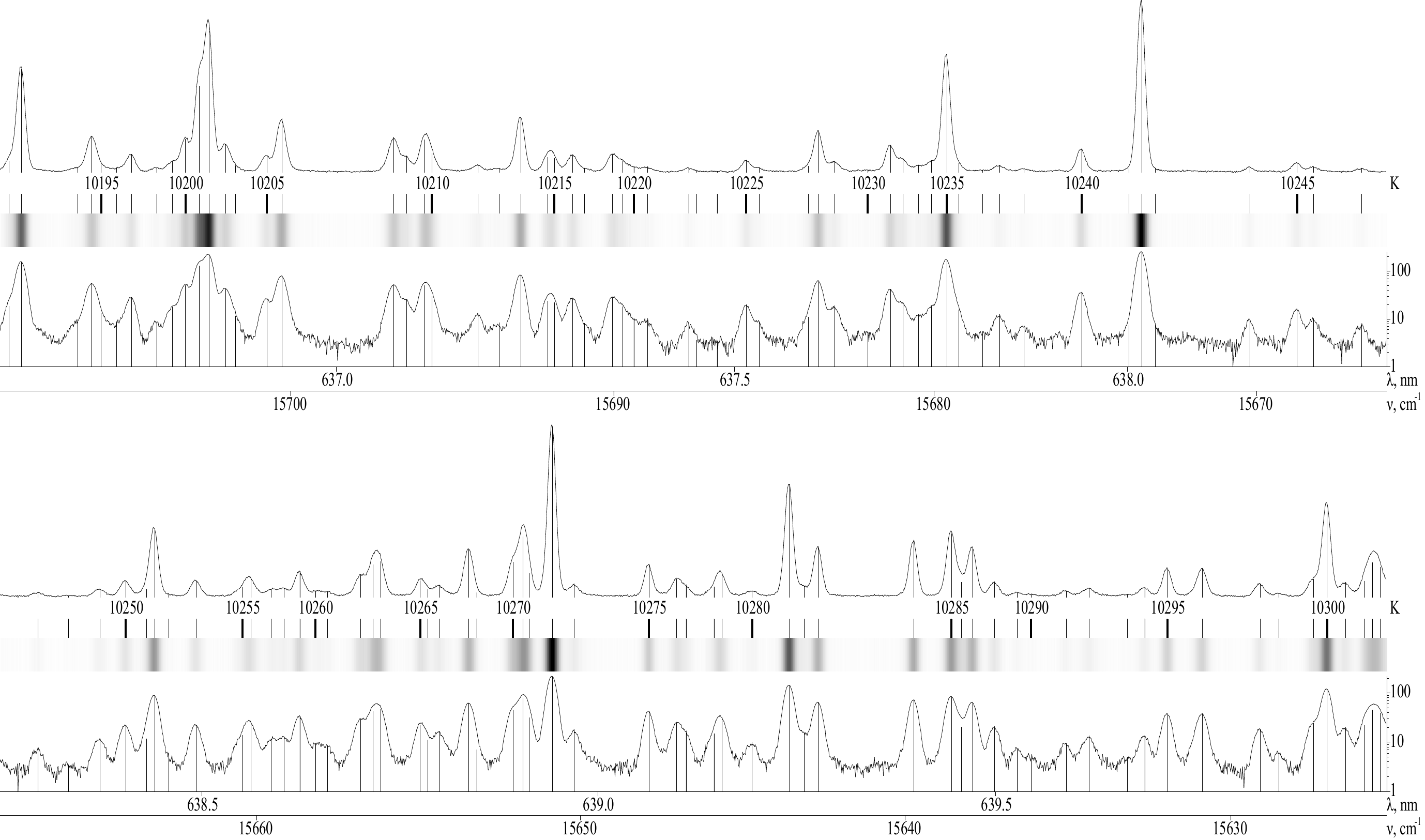}
\end{figure}

\newpage
\begin{figure}[!ht]
\includegraphics[angle=90, totalheight=0.9\textheight]{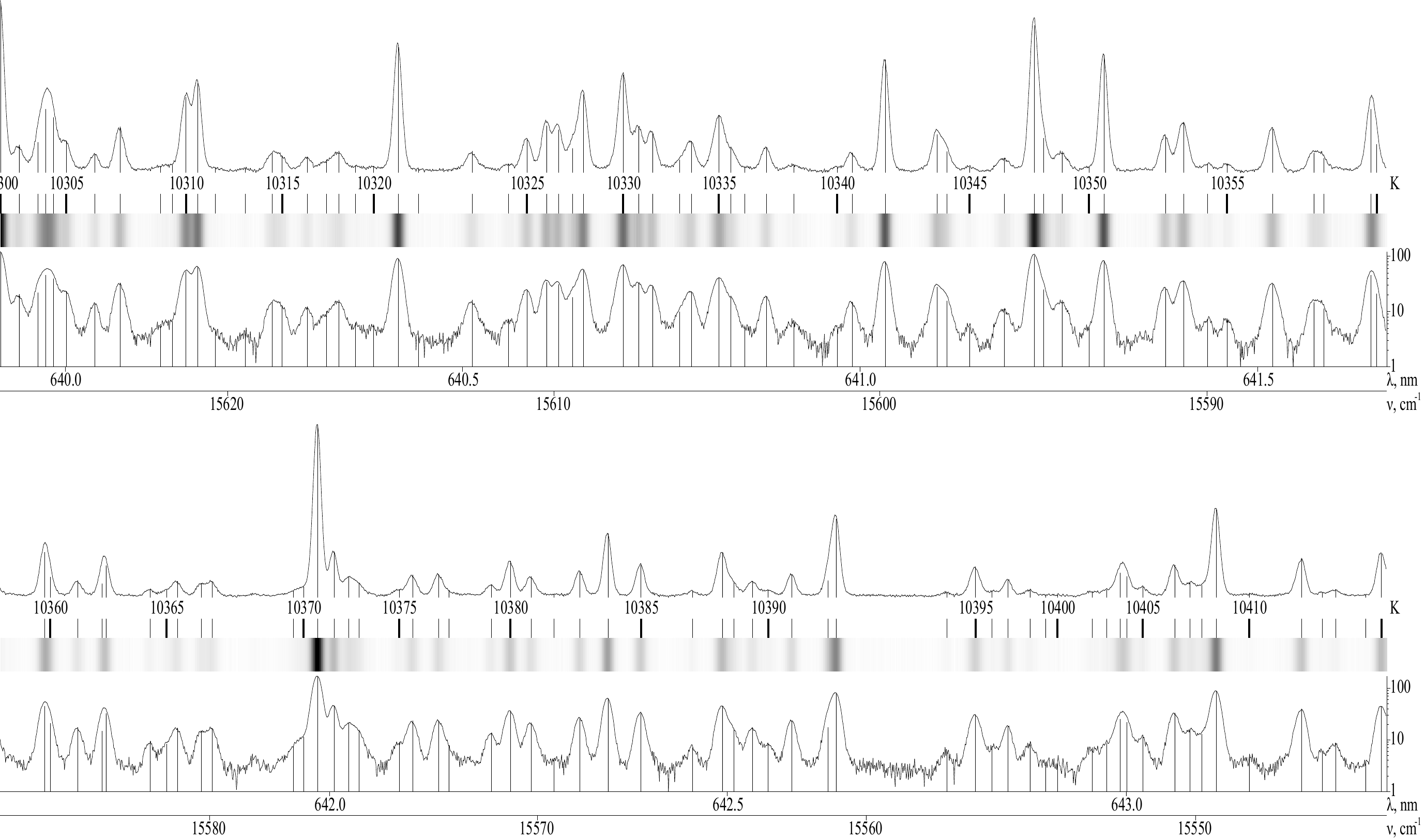}
\end{figure}

\newpage
\begin{figure}[!ht]
\includegraphics[angle=90, totalheight=0.9\textheight]{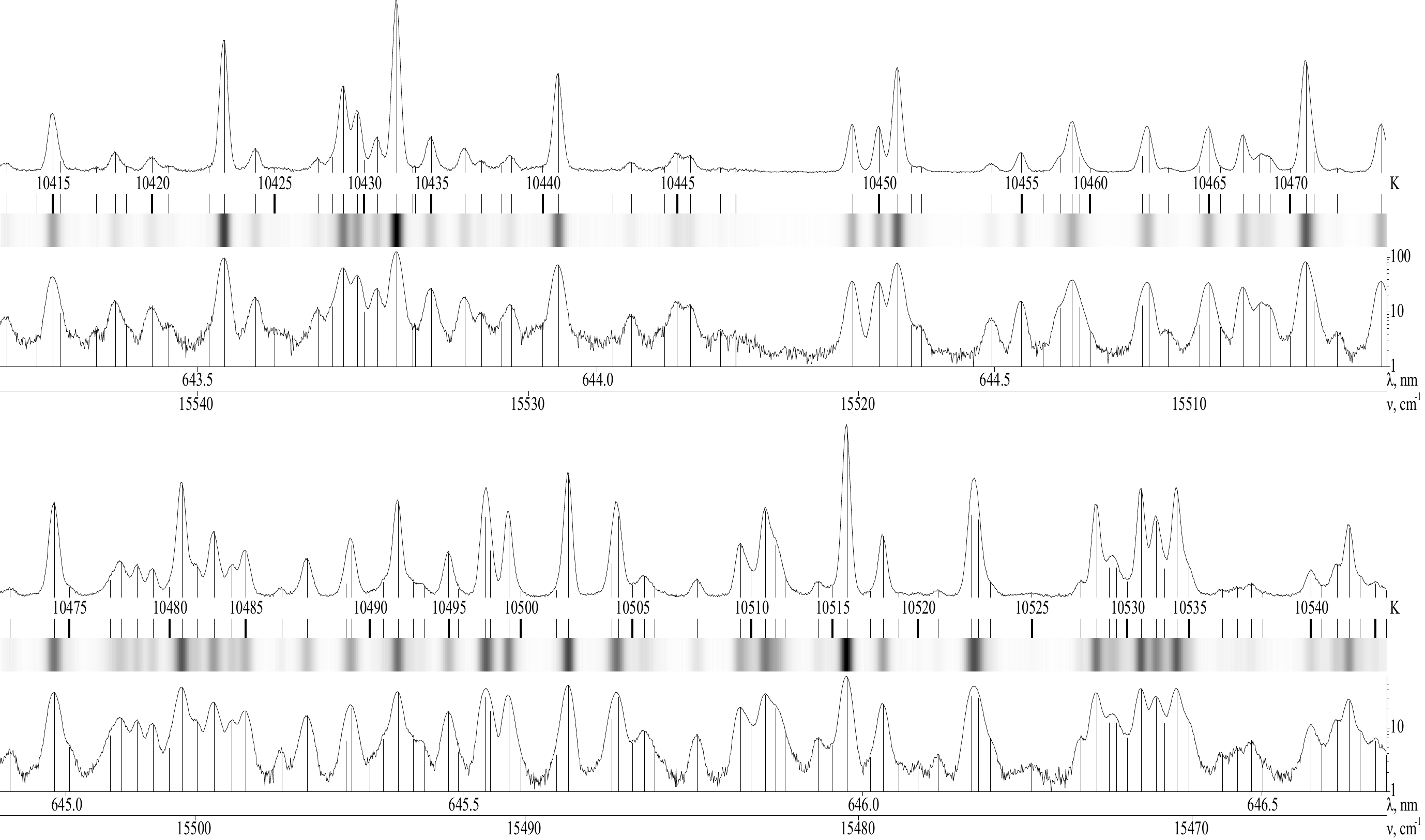}
\end{figure}

\newpage
\begin{figure}[!ht]
\includegraphics[angle=90, totalheight=0.9\textheight]{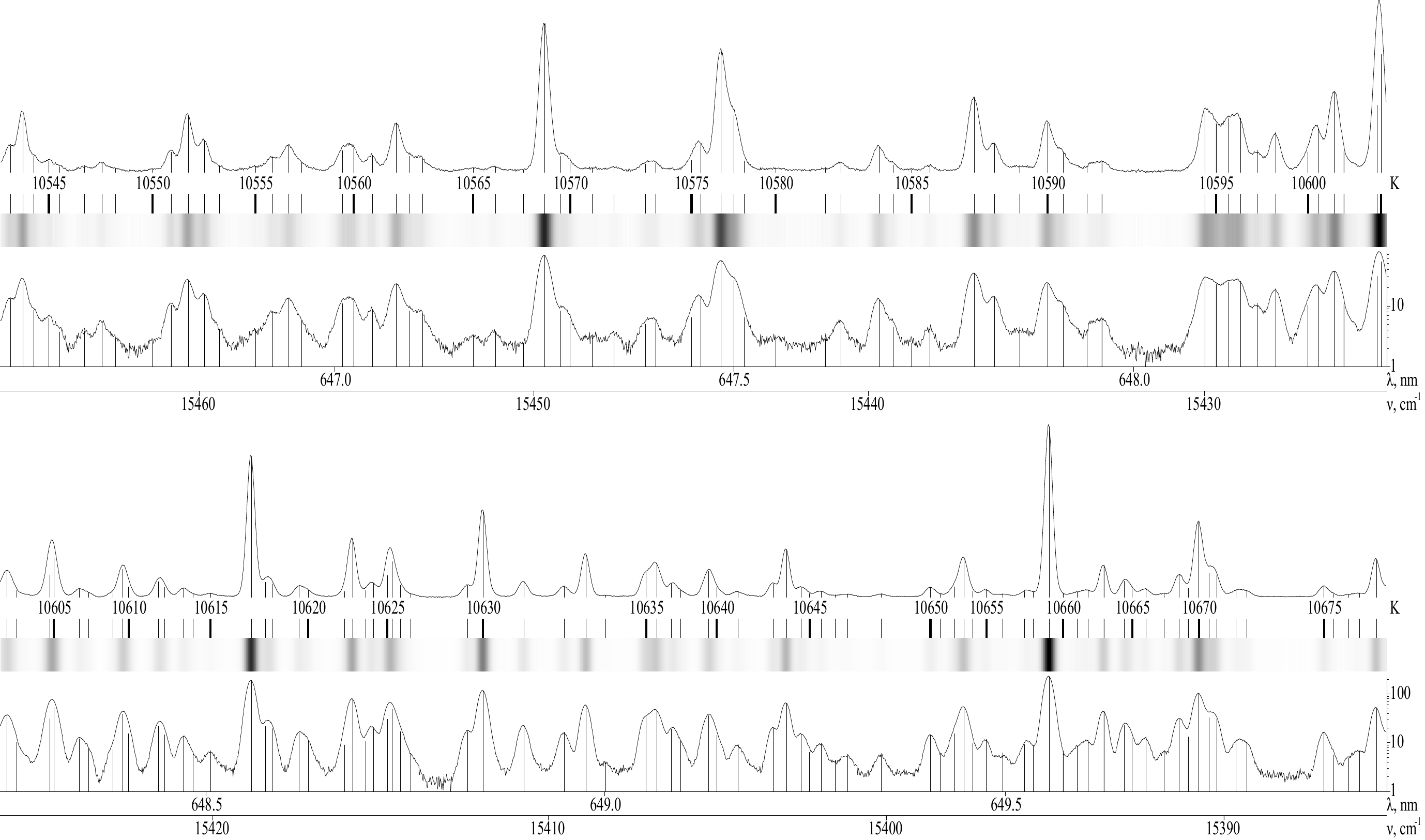}
\end{figure}

\newpage
\begin{figure}[!ht]
\includegraphics[angle=90, totalheight=0.9\textheight]{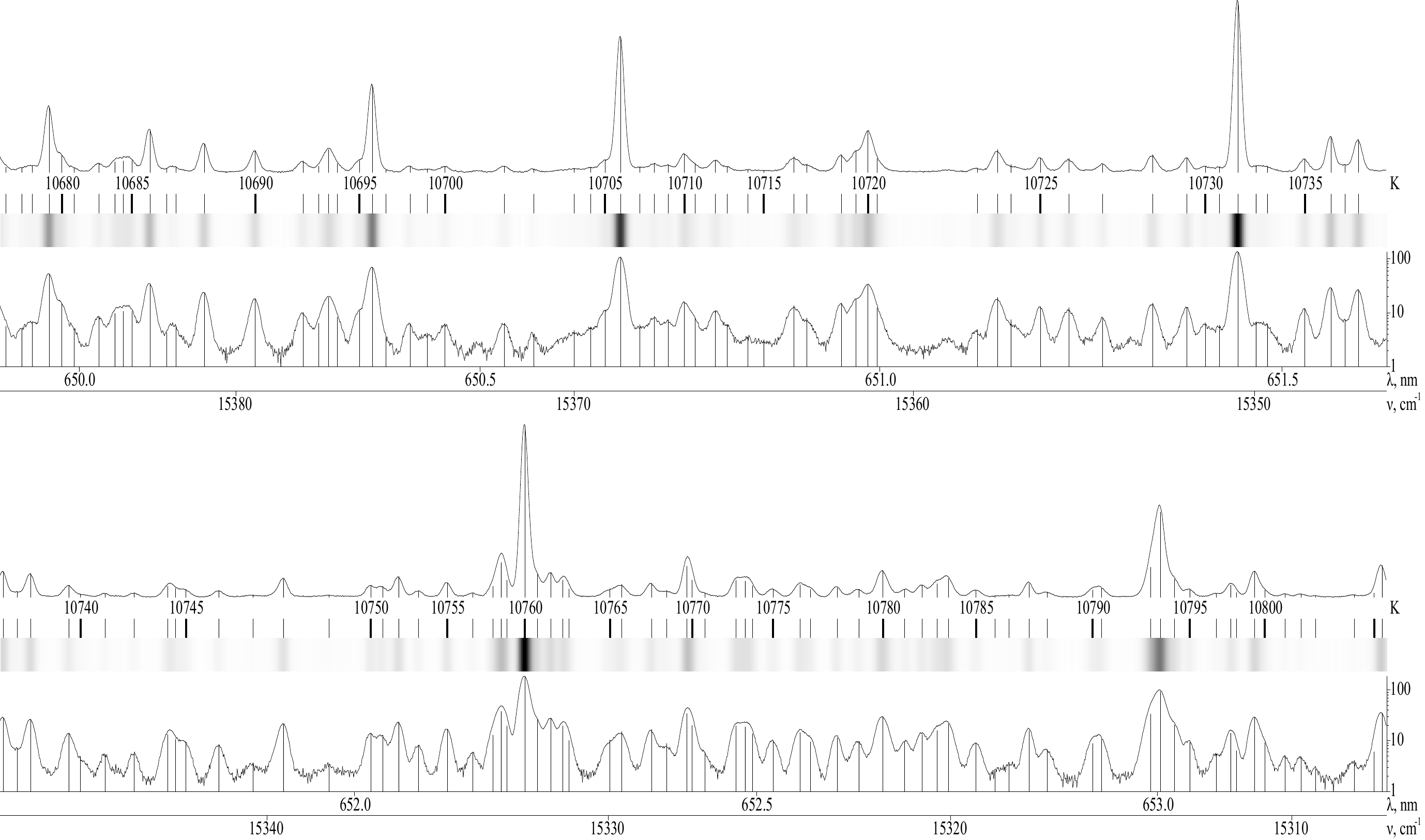}
\end{figure}

\newpage
\begin{figure}[!ht]
\includegraphics[angle=90, totalheight=0.9\textheight]{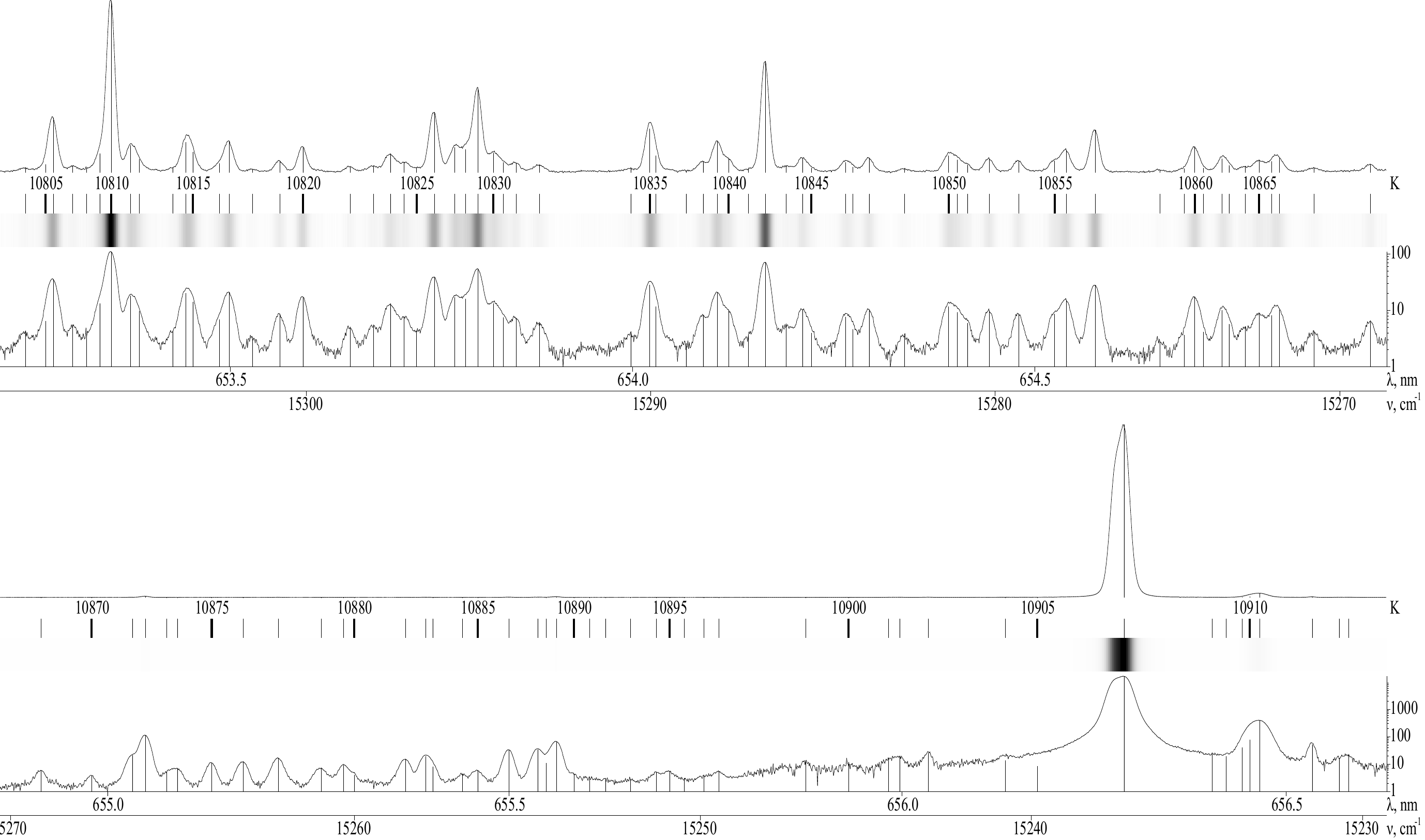}
\end{figure}

\newpage
\begin{figure}[!ht]
\includegraphics[angle=90, totalheight=0.9\textheight]{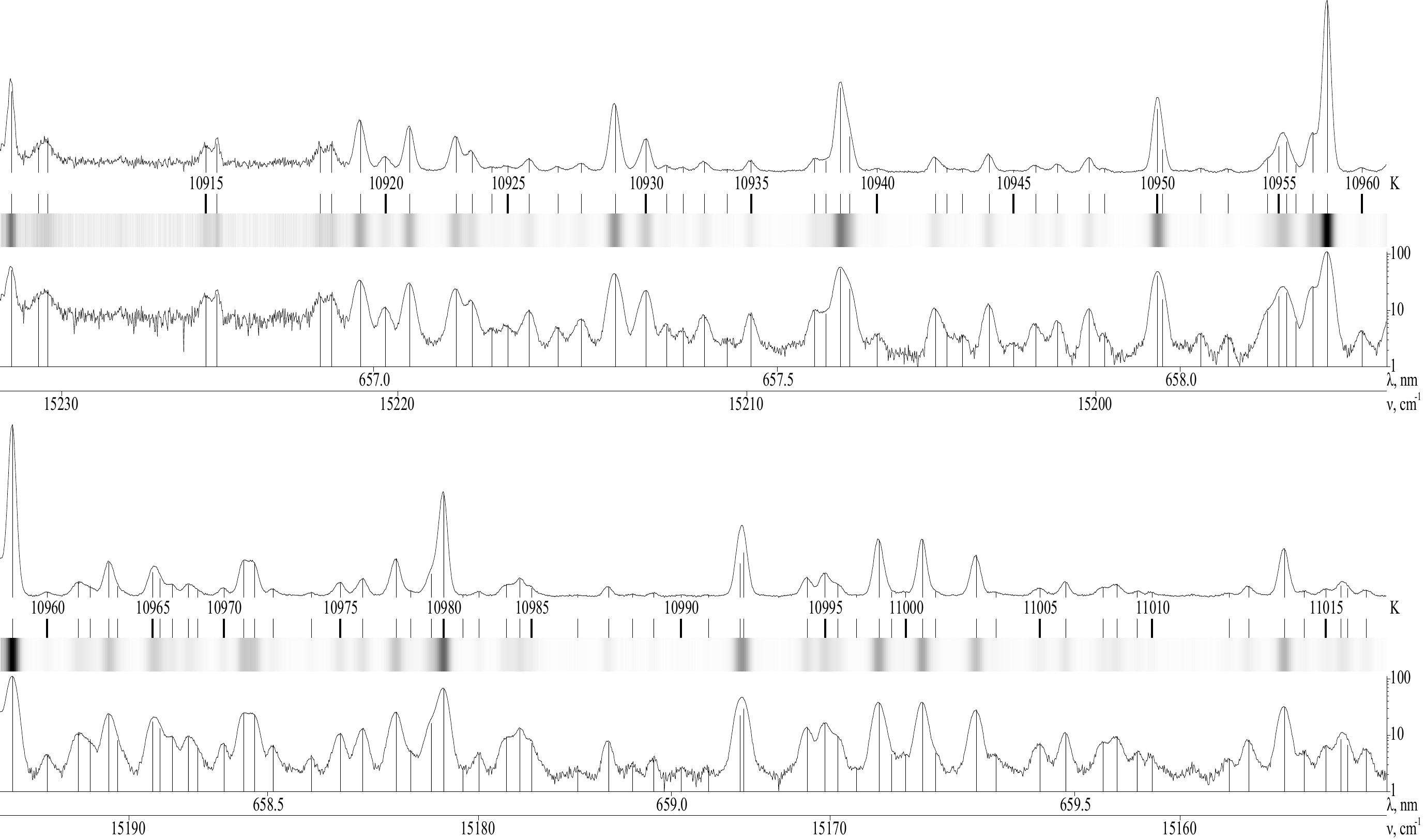}
\end{figure}

\newpage
\begin{figure}[!ht]
\includegraphics[angle=90, totalheight=0.9\textheight]{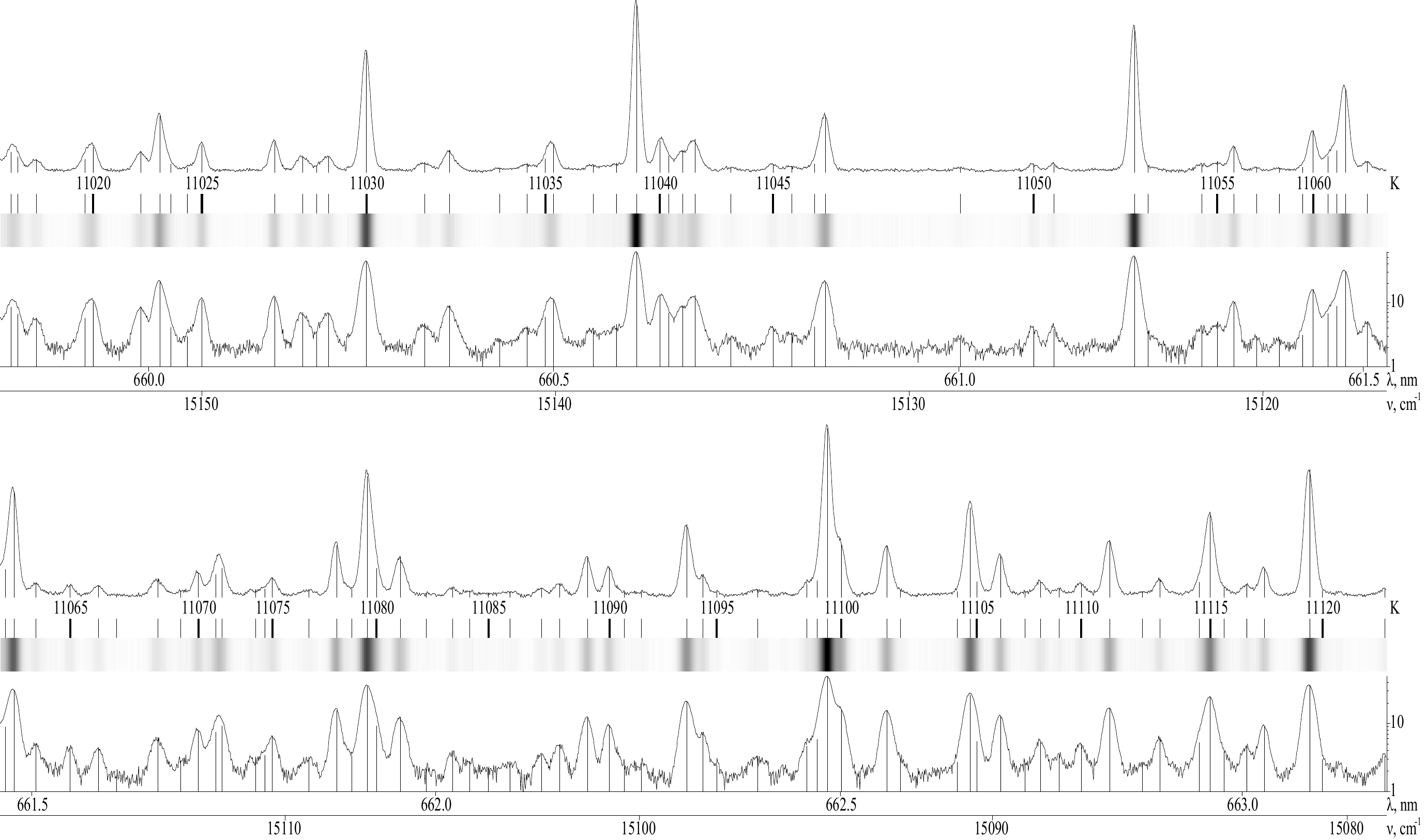}
\end{figure}

\newpage
\begin{figure}[!ht]
\includegraphics[angle=90, totalheight=0.9\textheight]{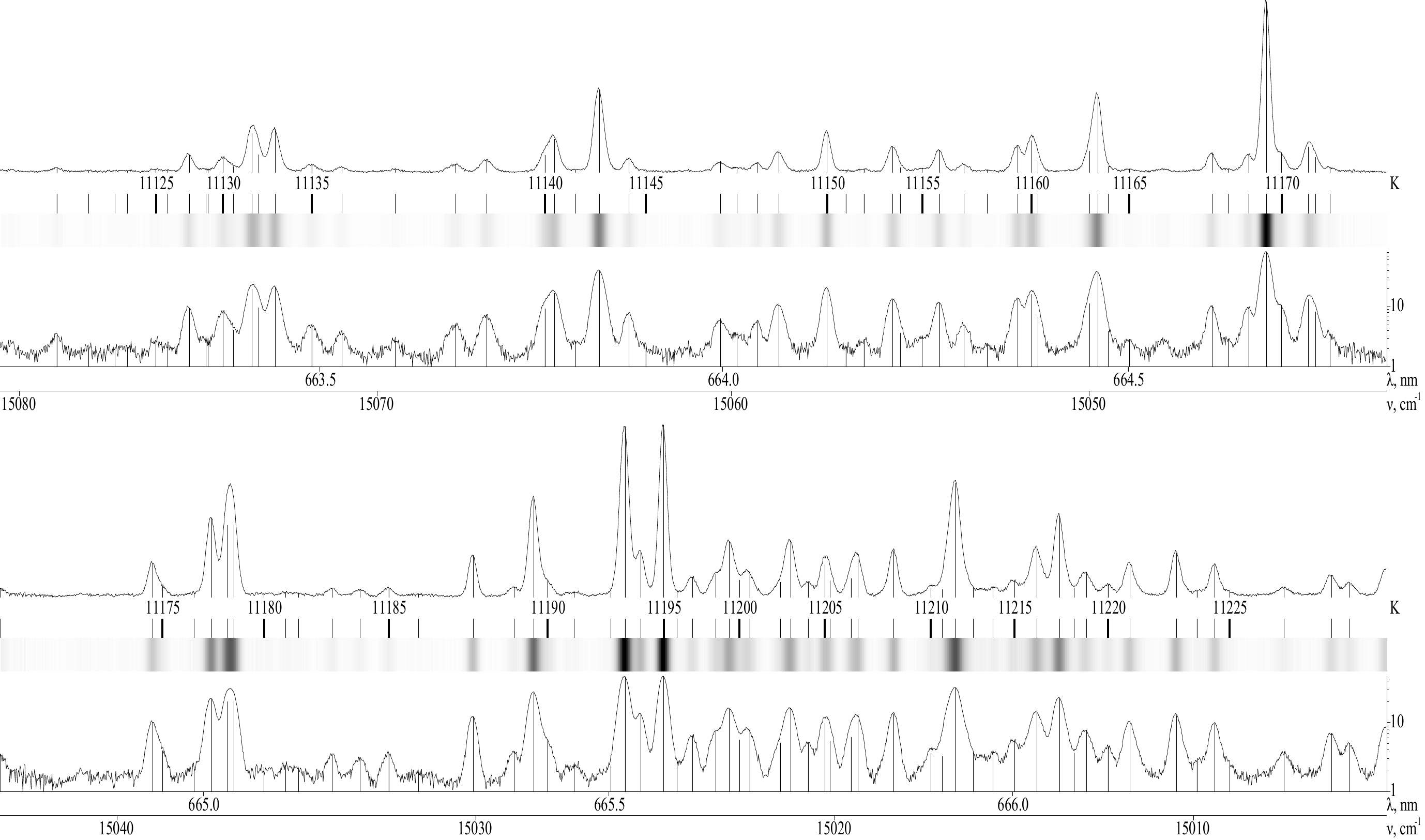}
\end{figure}

\newpage
\begin{figure}[!ht]
\includegraphics[angle=90, totalheight=0.9\textheight]{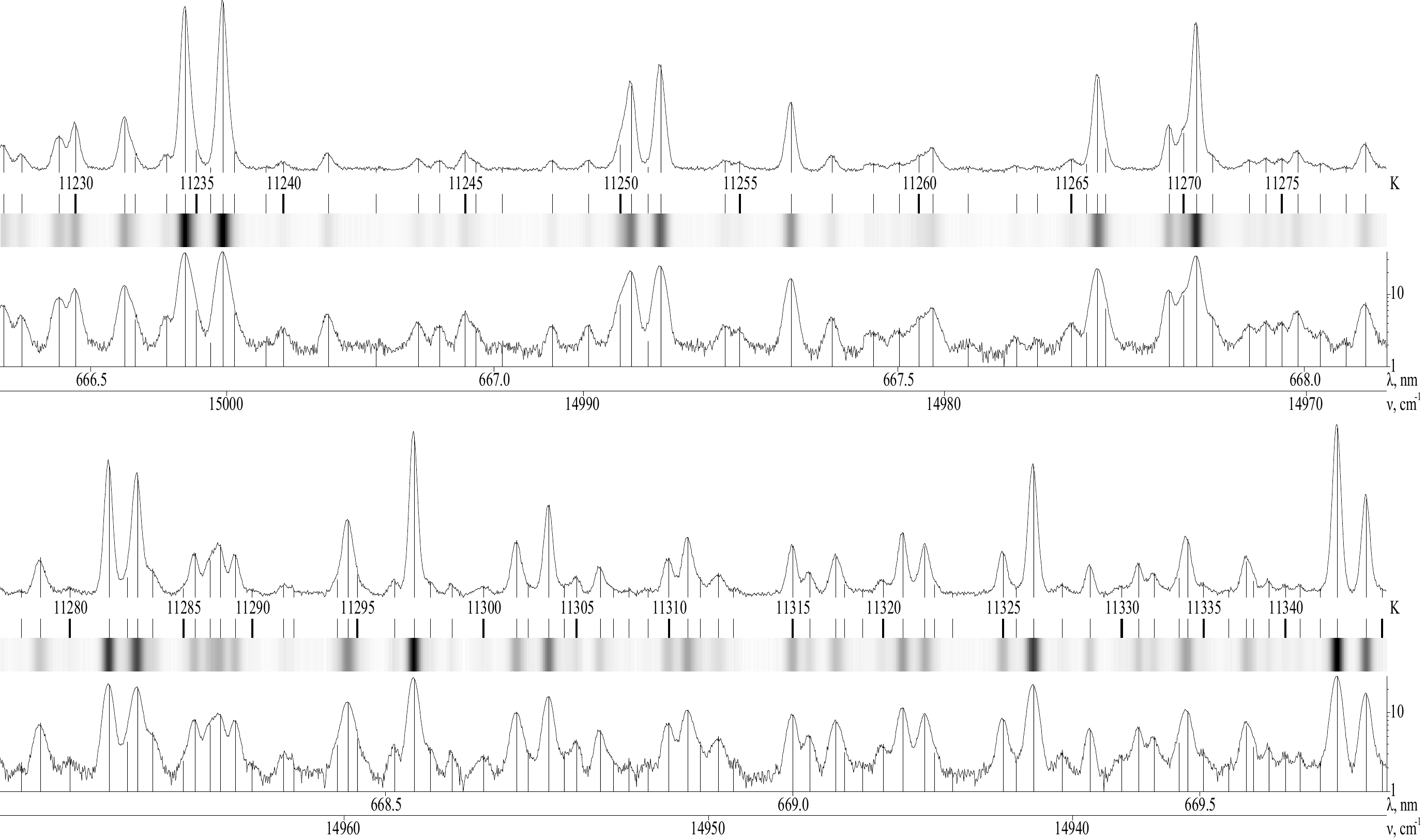}
\end{figure}

\newpage
\begin{figure}[!ht]
\includegraphics[angle=90, totalheight=0.9\textheight]{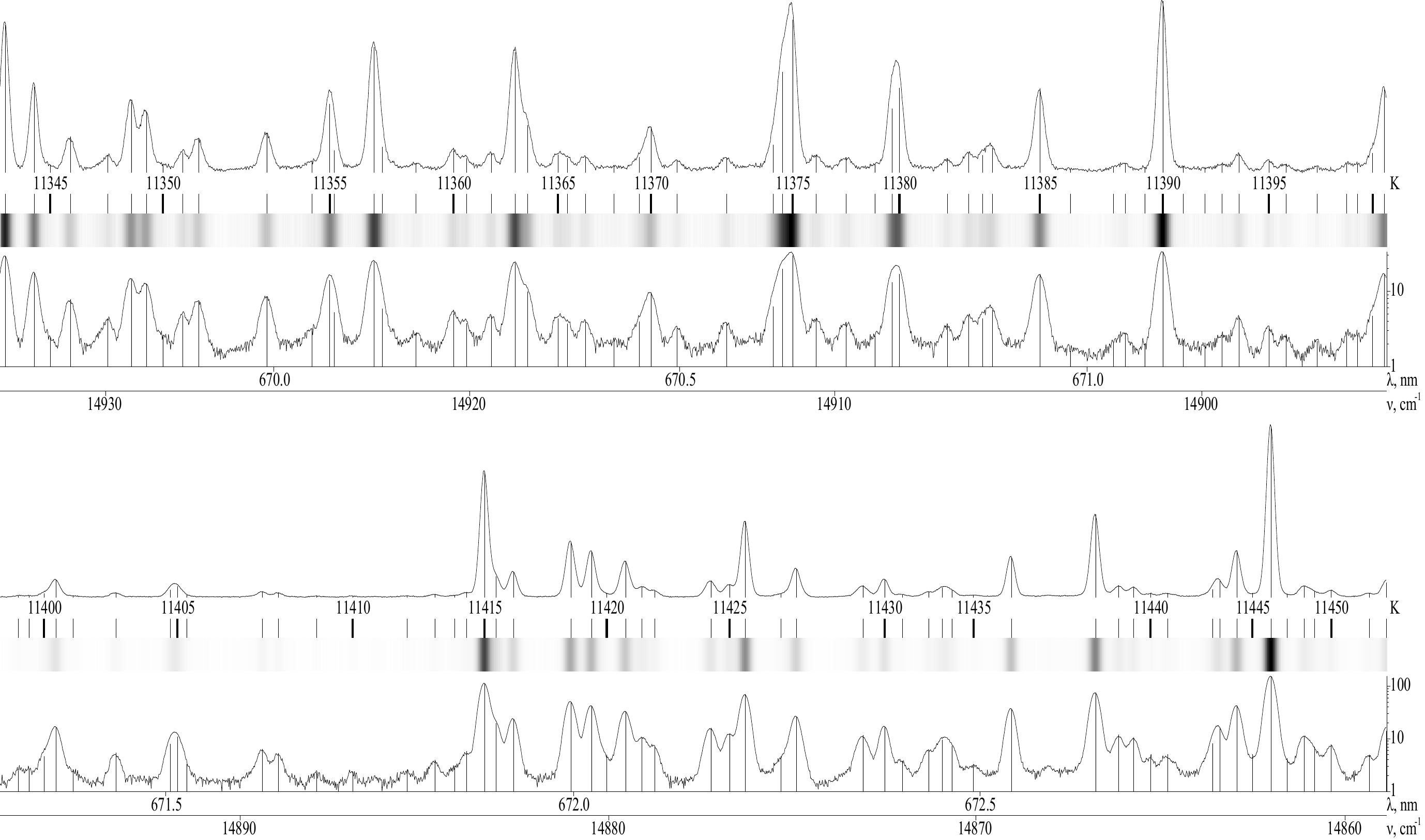}
\end{figure}

\newpage
\begin{figure}[!ht]
\includegraphics[angle=90, totalheight=0.9\textheight]{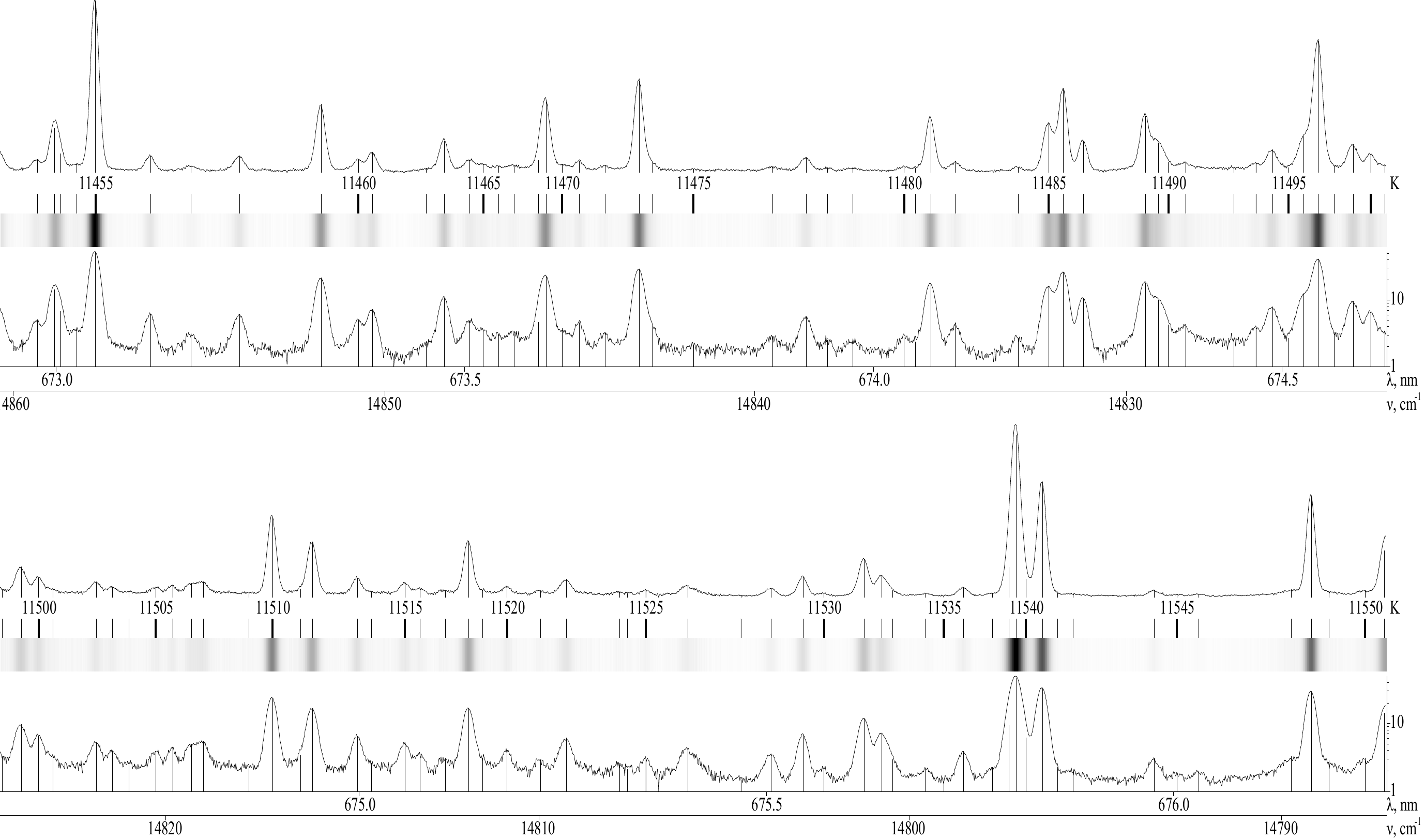}
\end{figure}

\newpage
\begin{figure}[!ht]
\includegraphics[angle=90, totalheight=0.9\textheight]{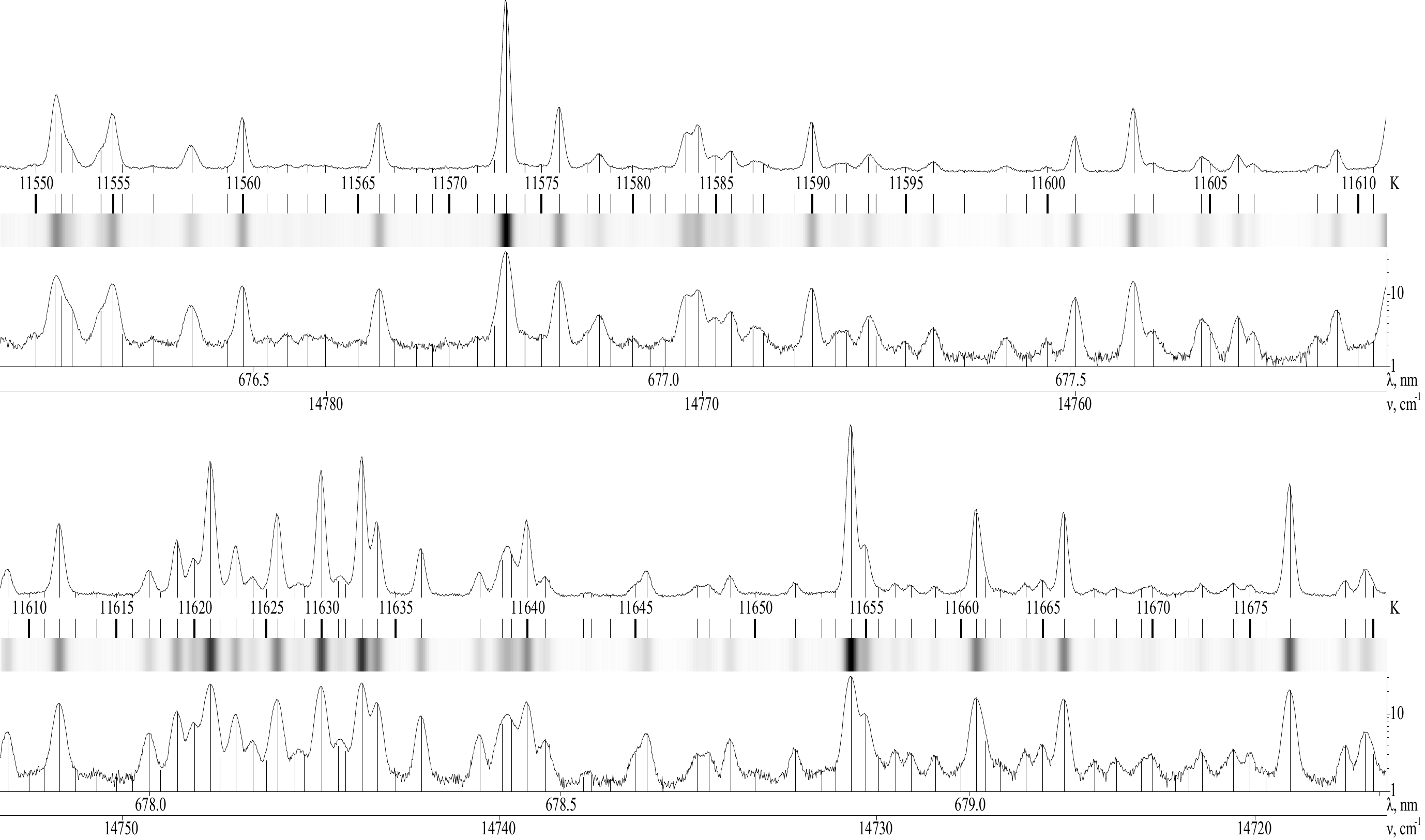}
\end{figure}

\newpage
\begin{figure}[!ht]
\includegraphics[angle=90, totalheight=0.9\textheight]{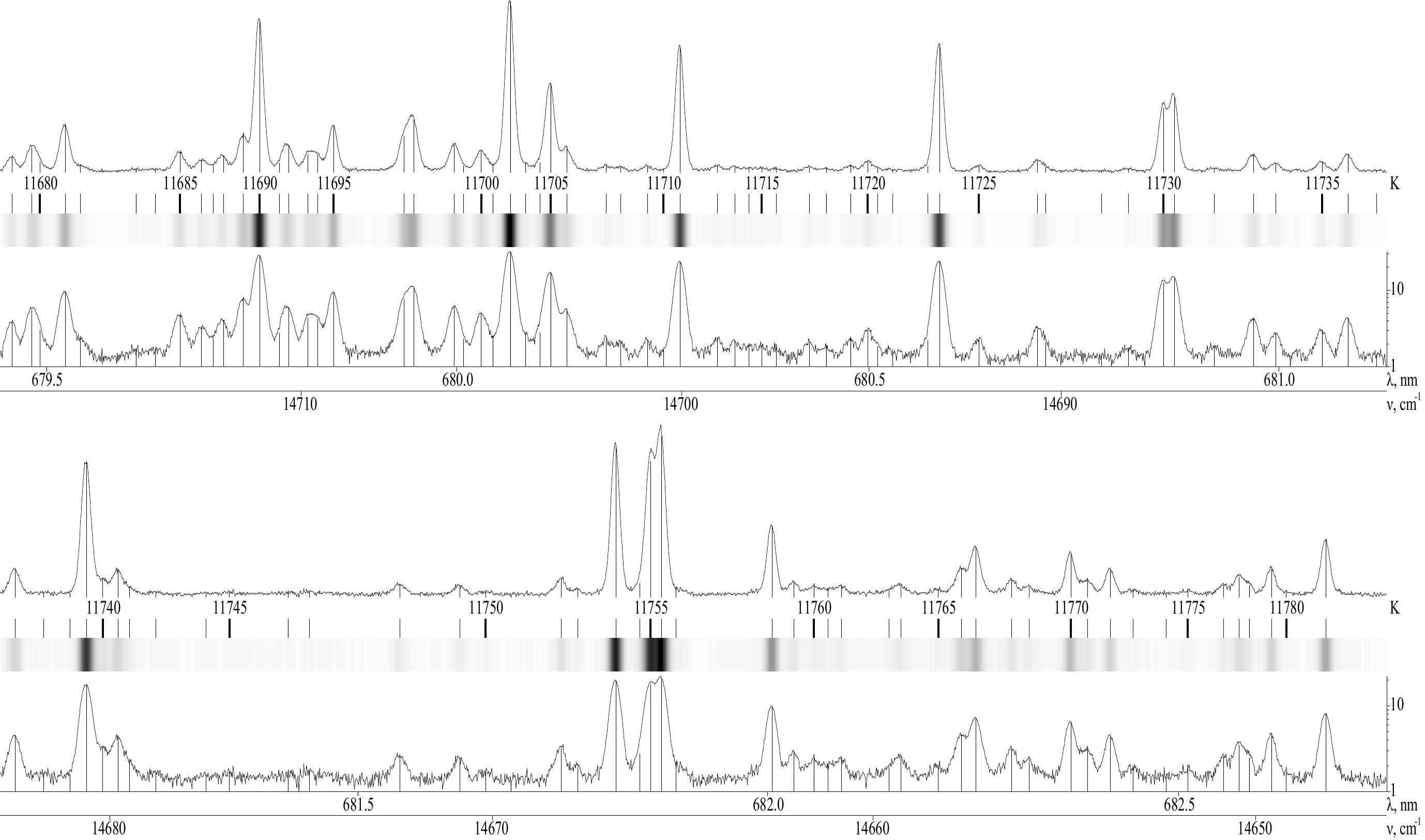}
\end{figure}

\newpage
\begin{figure}[!ht]
\includegraphics[angle=90, totalheight=0.9\textheight]{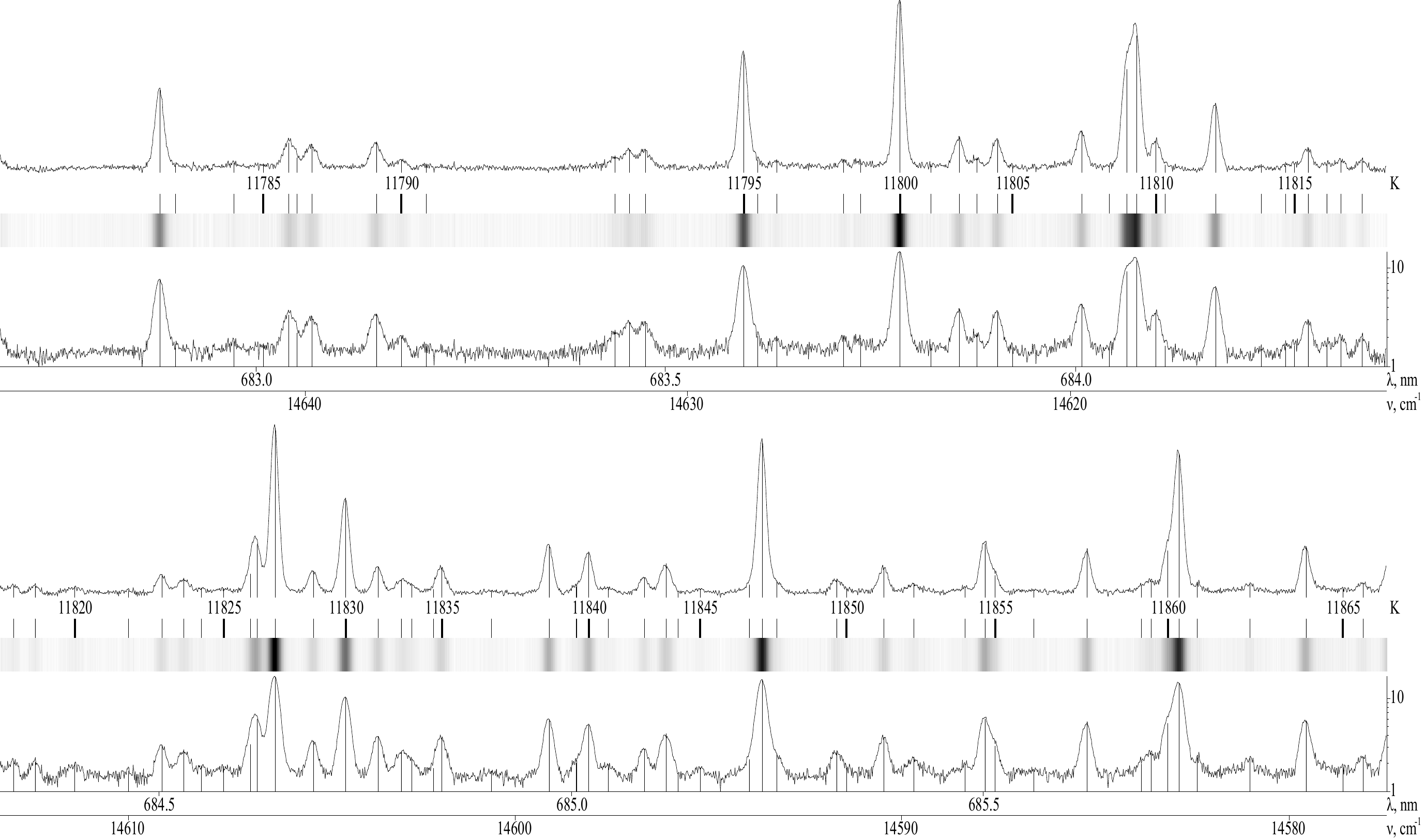}
\end{figure}

\newpage
\begin{figure}[!ht]
\includegraphics[angle=90, totalheight=0.9\textheight]{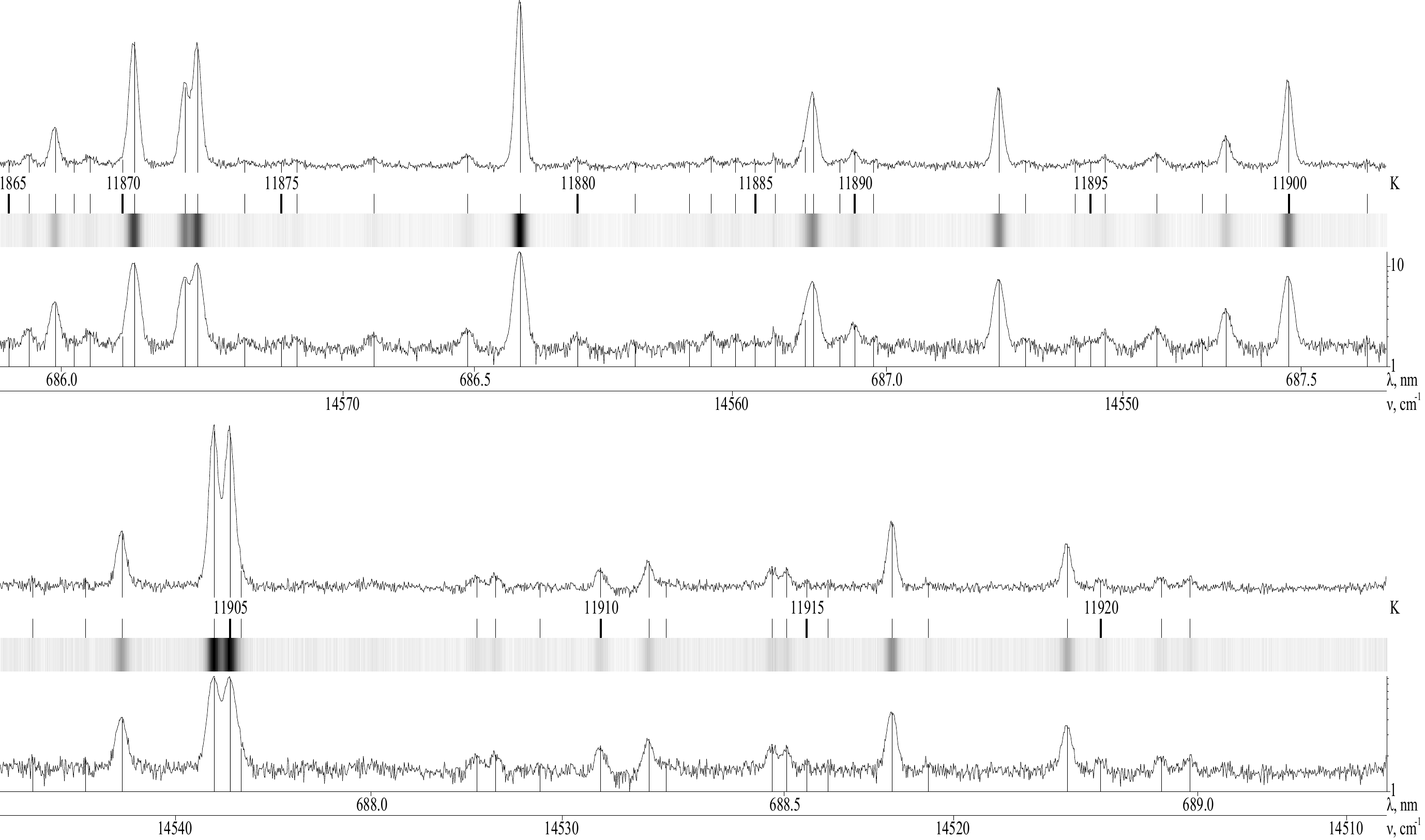}
\end{figure}

\newpage
\begin{figure}[!ht]
\includegraphics[angle=90, totalheight=0.9\textheight]{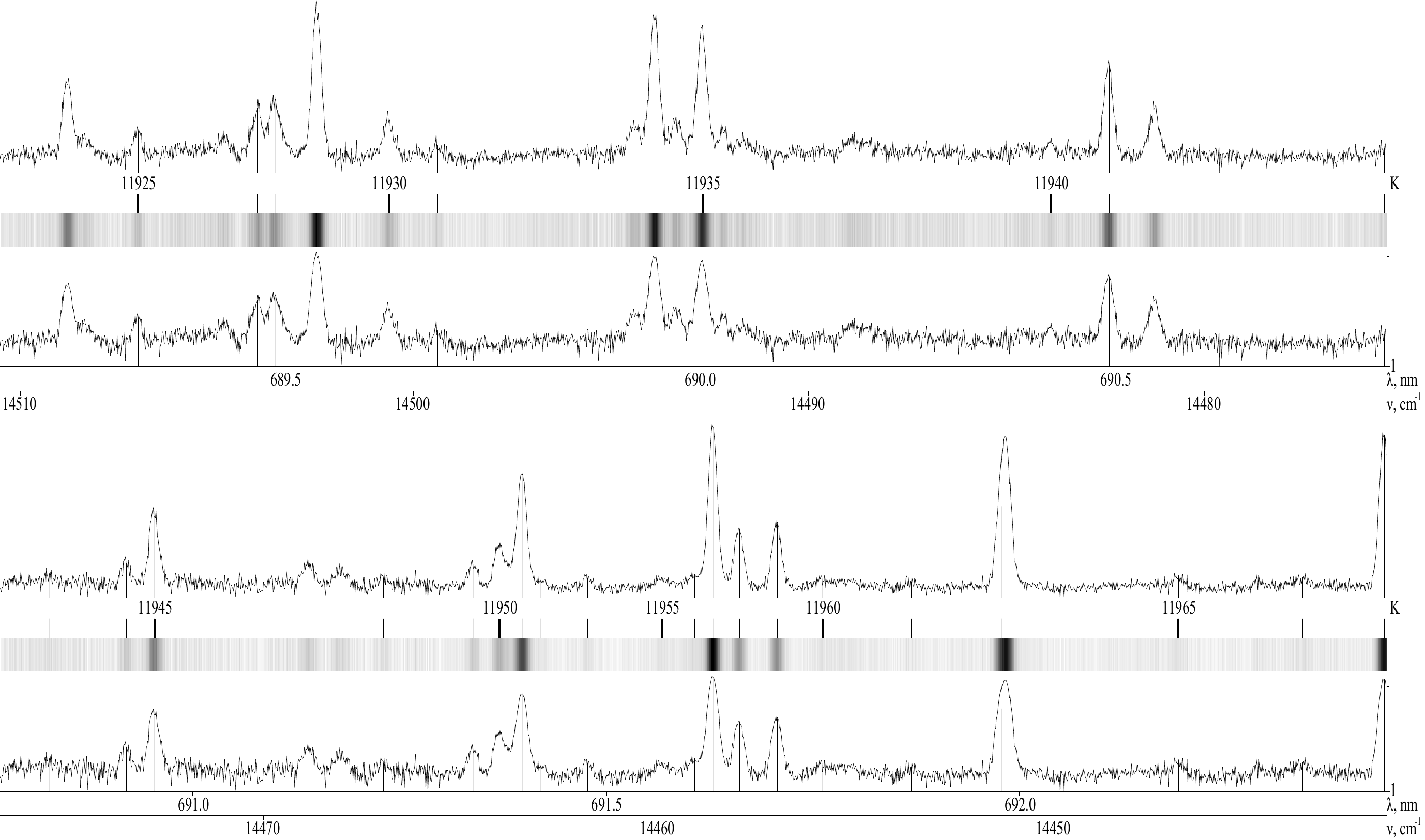}
\end{figure}

\newpage
\begin{figure}[!ht]
\includegraphics[angle=90, totalheight=0.9\textheight]{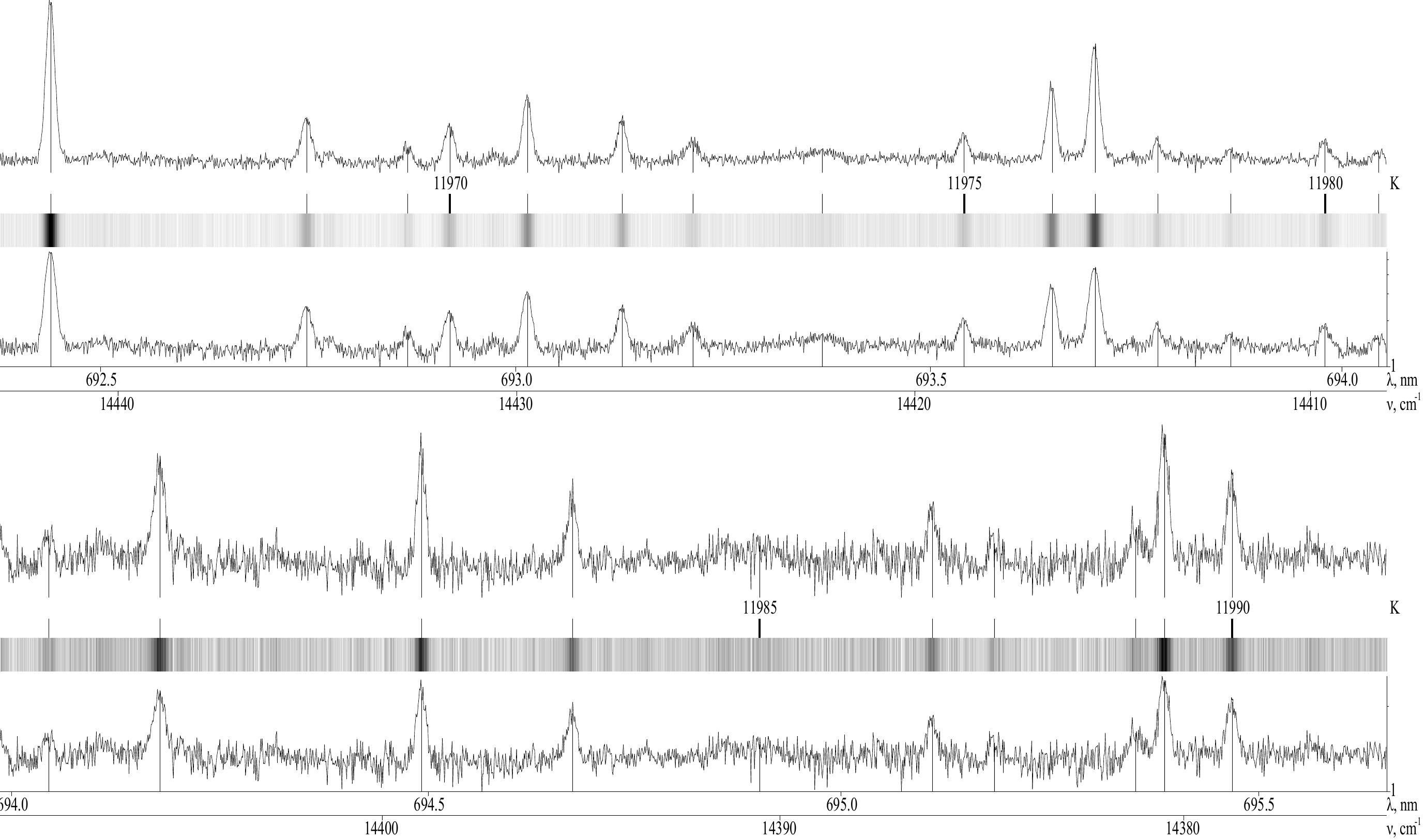}
\end{figure}

\end{document}